\documentclass[
reprint,
 amsmath,amssymb,
 aps,
]{revtex4-2}

\usepackage{graphicx}
\usepackage{dcolumn}
\usepackage{bm}
\usepackage{amssymb}
\usepackage{amsmath, amsthm}
\usepackage{mathrsfs}
\usepackage{supertabular, multirow}
\usepackage{graphicx}
\usepackage{epsfig}
\usepackage{subfigure}
\usepackage{bm}
\usepackage{xcolor}
\usepackage{algorithm}
\usepackage{tabularx}

\newtheorem*{theorem*}{Theorem}

\newcommand{\vsLt}{L^2}

\newcommand{\oC}{\mathcal C}

\begin{document}

\title{Reducing data resolution for better super-resolution:\\ {Reconstructing} turbulent flows from noisy observation}

\author{Kyongmin Yeo}
\email[Author to whom all correspondence should be addressed: ]{kyeo@us.ibm.com}
\affiliation{%
 IBM T.J. Watson Research Center, NY, USA
}%

\author{Ma{\l}gorzata J. Zimo\'{n}}
\thanks{M.J.Z contributed equally to this work with K.Y.}
\affiliation{IBM Research - Europe, UK
}
\altaffiliation[Also at ]{Department of Mathematics, University of Manchester, UK}

\author{Mykhaylo Zayats}
\author{Sergiy Zhuk}
\affiliation{
 IBM Research - Europe, Ireland
}

\date{\today}

\begin{abstract}
A super-resolution (SR) method for the reconstruction of Navier-Stokes (NS) flows from noisy observations is presented. In the SR method, first the observation data is averaged over a coarse grid to reduce the noise at the expense of losing resolution and, then, a dynamic observer is employed to reconstruct the flow field by reversing back the lost information. We provide a theoretical analysis, which indicates a chaos synchronization of the SR observer with the reference NS flow. It is shown that, even with noisy observations, the SR observer converges toward the reference NS flow exponentially fast, and the deviation of the observer from the reference system is bounded. Counter-intuitively, our theoretical analysis shows that the deviation can be reduced by increasing the lengthscale of the spatial average, i.e., making the resolution coarser. The theoretical analysis is confirmed by numerical experiments of two-dimensional NS flows. The numerical experiments suggest that there is a critical lengthscale for the spatial average, below which making the resolution coarser improves the reconstruction.
\end{abstract}

\maketitle




Super-resolution (SR) refers to the problem of reconstructing high-resolution information from low-resolution data~\cite{Capel03,Farsiu04}. It has been extensively studied in computer vision \cite{Anwar20} and image analysis~\cite{Wang19,Fang21}. Recently, with the advances in deep learning, there has been encouraging results in the applications of SR for physics problems~\cite{Banerjee23,Jiang20,Guemes22}, including complex turbulent flows~\cite{Fukami21,kim21,Zayats22}. 

Considering the chaotic nature of Navier-Stokes (NS) flows, SR reconstruction of flow field raises an interesting question. With the absence of the small-scale information, it is natural to assume that the chaotic dynamics will make the reconstructed flow field diverge from the ground truth, which will render the applicability of SR on flow reconstruction limited. Chaos synchronization, which refers to fascinating phenomena where multiple chaotic systems spontaneously synchronize with each other by exchanging information \cite{Pecora90,Pecora15}, provides a hint about how SR of the flow field is possible. The pioneering studies on the chaos synchronization in turbulent flows \cite{Yoshida05,Lalescu13} revealed that small-scale motions, \emph{e.g}, below about 20 Kolmogorov dissipation lengths, are slaved to the large-scale chaotic dynamics. Recently, \cite{Inubushi23} suggested that the chaos synchronization is an inherent property of NS equations. For two-dimensional NS flows, \cite{Zhuk23} mathematically proved the convergence of SR, i.e., synchronization of an observer system with the reference system through SR flow reconstruction, and proposed a critical lengthscale required for the convergence. It is important to note that most of the previous studies \cite{Yoshida05,Lalescu13,Leoni20,Zhuk23,Wang22,Inubushi23} consider the synchronization in the absence of a noise in the data. It is not well understood if the similar chaos synchronization will occur when the data from the reference is corrupted, which is typical in experiments \cite{Zaki21}.

In this study, we consider SR of two-dimensional NS flows with corrupted data. We show, both theoretically and in numerical experiments, that SR can lead to a better reconstruction of the flow state from highly noisy observations. Although deep-learning SR models have shown promising results \cite{Fukami23}, we employ a dynamic minimax estimator~\cite{Baras1995,Bardi1997} to facilitate theoretical analysis. This estimator is a partial differential equation (PDE) in the form of the Luenberger observer~\cite{Luenberger71}: it injects data as an additional forcing term into the underlying physics model, which is similar to the ``nudging'' in Data Assimilation~\cite{Leoni20}.
Our theoretical analysis reveals that, while the observer may not be completely synchronized with the reference system due to the corrupted data, the deviation of the observer from the reference system is bounded and, surprisingly,  SR can reduce the size of the bounding ball.

\begin{figure*}[t]
    \centering
    \includegraphics[width=.95\textwidth]{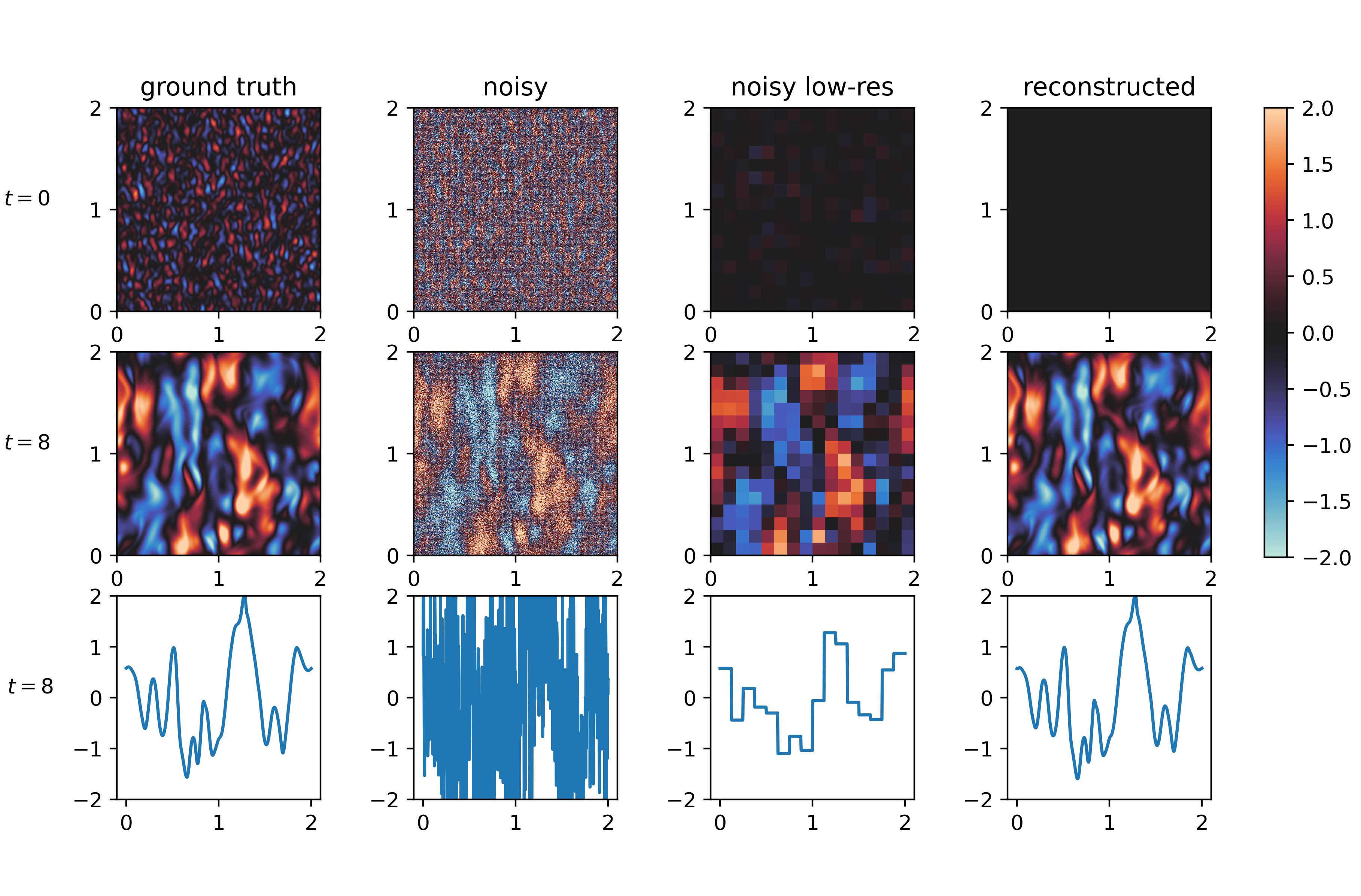}
    \caption{Snapshots of the $x_1$-component of the velocity for $\nu=0.0015$. From the left to the right columns, the plots denote the ground truth $u_1(\bm{x},t)$, noisy observation $y_1(\bm{x},t)$, low-resolution observation, $(\mathcal{C}y_1)(\bm{x},t)$, and the estimated state, $z_1(\bm{x},t)$. The low-resolution observation and the state estimation are for $c=32$, i.e., the subdomain size of $h=32 \delta_x$. The noise level is $\zeta = 1.6$, indicating the standard deviation of the noise is 1.6 times larger than TKE. The top row shows the initial condition and the middle row denotes the velocity at $t=8$. The bottom row shows the velocity profiles at the middle of the domain, i.e., $u_1(x_1,\pi,t)$. In the plots, the spatial coordinate is scaled by $\pi$. }
    \label{fig:snapshot}
\end{figure*}

In our SR model, first the noisy observation is transformed into a lower resolution data by a linear operator, which reduces the noise, albeit at the expense of losing information. In particular, we employ a spatial averaging over a coarser grid given by non-overlapping coverings. Then, the Luenberger observer reverses back the lost information. 

Let $\bm{u}(\bm{x},t)$ be the solution of the two-dimensional incompressible Navier-Stokes equations (NSE) in a rectangular domain ($\Omega$) with periodic boundary conditions,  and $\bm{y}$ be a noisy observation of $\bm{u}$;
\begin{equation}
    \bm{y}(\bm{x},t) = \bm{u}(\bm{x},t)+\bm{\eta}(\bm{x},t).
\end{equation}
Here, $\bm{\eta}$ is a zero-mean white noise with a variance, $\sigma^2$. Note that we do not restrict $\bm{\eta}$ to be Gaussian, but $\bm{\eta}$ is assumed to satisfy a regularity condition (see Eq. 6 in Supplementary Material).

Assume that $\Omega$ can be partitioned into $N$ disjoint rectangles; $\Omega = \bigcup_{j=1}^N \Omega_j$. Define a sampling operator $\oC$ as
\begin{equation} 
(\mathcal{C}\bm{u})(\bm{x},t) = \sum_{j=1}^N \left(\frac{1}{|\Omega_j|}
    \int_{\Omega_j}\bm{u}(\bm{x},t)~ d\bm{x}\right) \xi_j(\bm{x}).\label{eq:Coverages}
\end{equation}
Here, $\xi_j(\bm{x})$ is an indicator function for $\Omega_j$, which is one if $\bm{x} \in \Omega_j$ and zero otherwise.
Then, a low-resolution observation is generated as
\[
\bm{Y} = \mathcal{C}\bm{y}.
\]
In other words, $\oC$ takes noisy high-resolution data, $\bm{y}$, and maps it to a low-resolution data, $\bm{Y}$, which is piece-wise constant in space. 

Let $\bm{z}(\bm{x},t)$ denote the unique solution of the Luenberger observer \cite{Zhuk23}: 
\begin{align}
&\frac{\partial \bm{z}}{\partial t} = -\frac{1}{\rho}\bm{\nabla}p -\bm{z}\cdot \bm{\nabla}\bm{z} + \nu\nabla^2\bm{z}+\bm{f}+L(\bm{Y}-\mathcal{C}\bm{z}), \label{eqn:observer} \\
&\bm{\nabla}\cdot\bm{z} = 0,\label{eqn:div_freez}
\end{align}
where $\bm{f}(\bm{x},t)$ is a known exogenous forcing, $p$ is the pressure and $\nu$ is the kinematic viscosity. The observer's structure has a copy of NSE and a special input $L(\bm{Y}-\mathcal{C}\bm{z})$ for a positive coefficient, $L>0$. This input enforces $\bm{z}$ to agree with the low-resolution data $\bm{Y}$. 



\begin{figure*}[t]
    \centering
    \includegraphics[width=.32\textwidth]{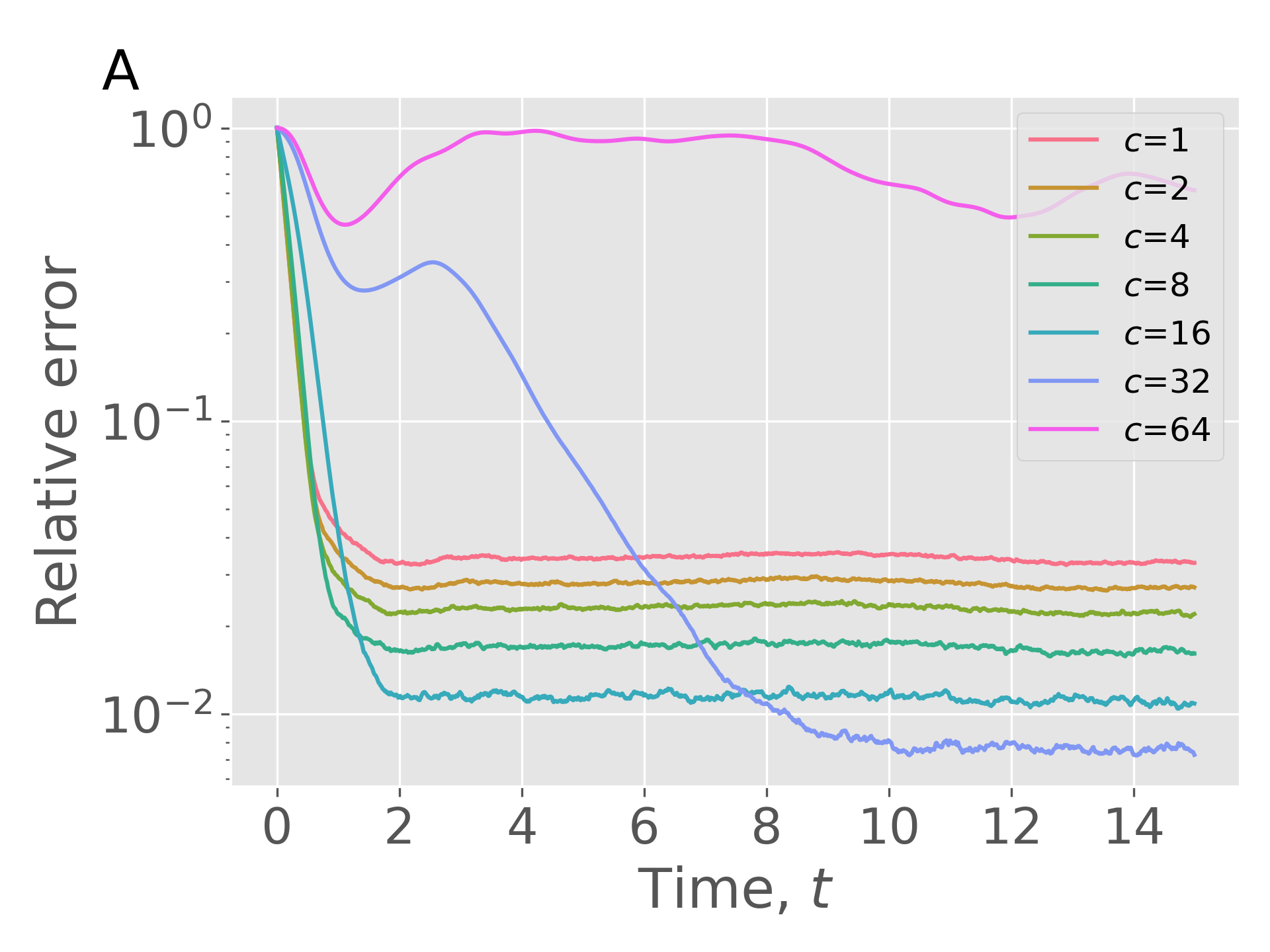}
    \includegraphics[width=.32\textwidth]{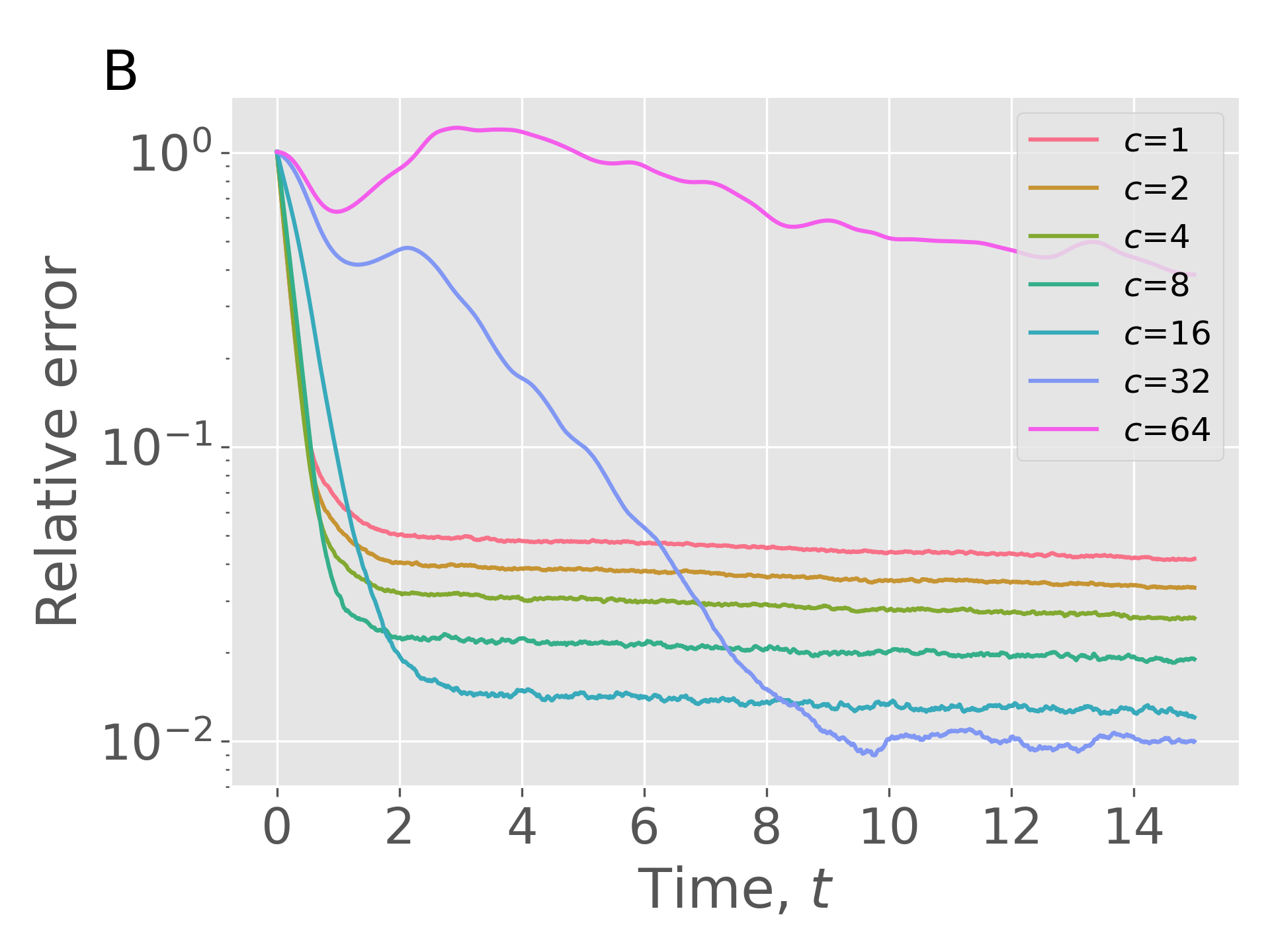}
    \includegraphics[width=.32\textwidth]{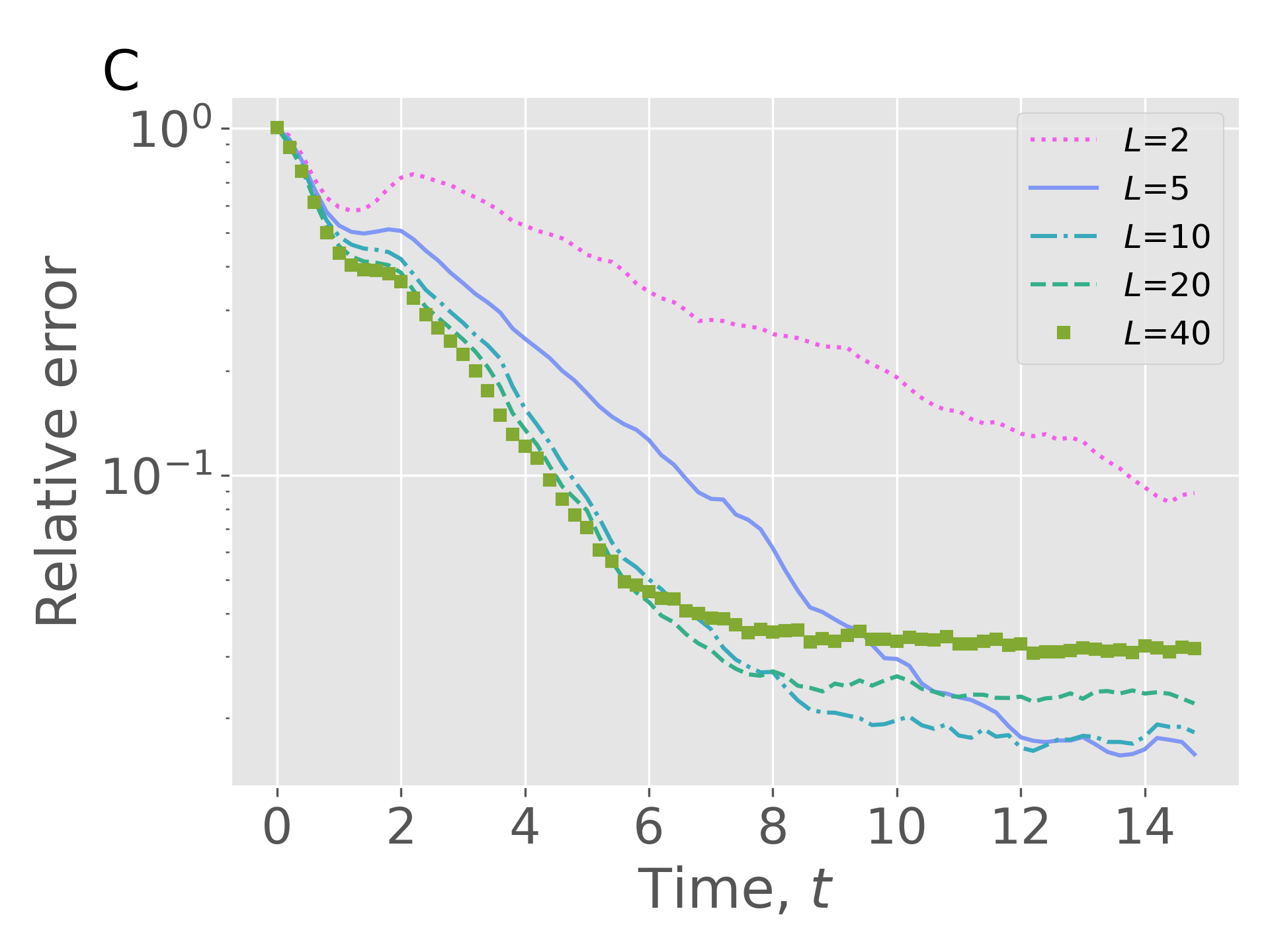}
    \caption{Fig. A and B show the temporal evolution of the error for a range of $c$ for $\nu=0.006$ and $0.003$, respectively. The noise magnitude is $\zeta=1.6$ and $L$ is set to 5. Fig. C shows the temporal changes of the error in terms of $L$ for $\nu=0.0015$, $c=32$ and $\zeta=1.6$.}
    \label{fig:time_error}
\end{figure*}

Let the estimation error be $\bm{e} = \bm{u}-\bm{z}$. Our theoretical analysis shows that there exists a convergence zone, $\Upsilon$, such that the evolution equation of $\bm{e}$ becomes
\begin{equation} \label{eqn:temporal_error}
    \frac{d \|\bm{e}\|^2_{\vsLt}}{dt} = -\gamma_1 - \gamma_2 \|\bm{e}\|^2_{\vsLt},
\end{equation}
when $\| \bm{e} \|_{L^2} > \Upsilon$. Here, $\| \cdot \|_{L^2}$ is an $L_2$ norm, and $\gamma_1 >0$ and $\gamma_2>0$ are problem-dependent constants. In other words, once $\| \bm{e} \|_{L^2}$  becomes larger than $\Upsilon$, $\bm{e}$ decays exponentially fast towards $\Upsilon$. Hence, $\Upsilon$ essentially provides an upper bound of the estimation error. Furthermore, if we assume that the noisy high-resolution data is obtained on a uniform grid with the grid resolution of $\delta_x$, the upper bound of the estimation error is
\begin{equation} \label{eqn:conv_zone}
\Upsilon = \sqrt{2} |\Omega|^{1/2} \frac{\sigma}{\delta c}.    
\end{equation}
Here, $\delta$ is a problem dependent constant and $c$ is a \emph{compression ratio}, i.e., ratio between the size of $\Omega_j$ and $\delta_x$; $c=|\Omega_j|^{1/2}/\delta_x$. Our theoretical analysis indicates that, counter-intuitively, the estimation error can be reduced by increasing the compression ratio, i.e., further reducing the resolution. The proof and derivation of the behavior of $\Upsilon$ are given in Supplementary Material.


To confirm our theoretical results, we performed a set of numerical simulations of two-dimensional turbulent flows.
The kinematic viscosity of the simulations is changed from $\nu=0.0015$ to 0.0075. Let the variance of the observation noise be
\begin{equation}\label{eq:noise_var}
    \sigma^2 = \zeta^2\times\text{TKE},    
\end{equation}
in which TKE denotes the turbulent kinetic energy
\begin{equation}\label{eq:tke}
    \text{TKE} = \frac{1}{2 |\Omega| } \| \bm{u} \|^2_{L^2}.    
\end{equation}
For each $\nu$, $\zeta$ is changed from 0.1 to 1.6 and the size of $\Omega_j$ is varied by changing $c=1$ to 64. Here, $c=1$ corresponds to estimating $\bm{u}$ directly from a noisy high-resolution observation, and $c>1$ indicates SR.

In the numerical experiments, the Navier-Stokes equations are solved in a two-dimensional rectangular domain, $\Omega=(0,2\pi)\times(0,2\pi)$ with the computational grid of size $512 \times 512$, which makes the grid resolution $\delta_x = 2\pi/512$. The observations are generated by a pseudo-spectral method \cite{spectral}. After computing $\bm{u}$, noisy observations, $\bm{y}$, are generated by adding a Gaussian noise. The state estimation is performed by solving the Luenberger observer, (\ref{eqn:observer}--\ref{eqn:div_freez}), using a second-order Finite Element method. Note that we use different solvers for the data generation, a highly accurate spectral method, and for the state estimation, a low-order finite element method, to mimic a realistic setting. {\color{black} The source code to reproduce the numerical experiments is publicly available \cite{our_software}.}

Fig.~\ref{fig:snapshot} displays snapshots of the $x_1$-component of the velocity ($u_1$) for $\nu=0.003$. The noisy observations, shown in the second column, are generated by adding Gaussian white noise of which $\sigma = 1.6 \text{TKE}$. Comparing $u_1(\bm{x},t)$ and $y_1(\bm{x},t)$, in particular in the bottom rows, it is shown that $\sigma$ is on the same order of magnitude with $u_1$, which makes it challenging to identify $u_1$. 

Our SR model aims to reconstruct $\bm{u}$ (the first column of Fig. \ref{fig:snapshot}) given $\bm{y}$ (the second column of Fig. \ref{fig:snapshot}). In this example, the low resolution data, $\bm{Y}$, is made by $|\Omega_i| = (32 \delta_x)\times(32 \delta_x)$. In other words, $\bm{y}$ of size $512 \times 512$, is downscaled to $16 \times 16$ using the observation operator ($\mathcal{C}$). The noise-filtering effect of $\mathcal{C}$ is clearly visible by comparing the second and third columns of Fig.~\ref{fig:snapshot}. At the same time, comparing the first and third columns, $\bm{Y}$ looks very different from $\bm{u}$ as small-scale eddies are averaged out.

The fourth column of the first row of Fig.~\ref{fig:snapshot} shows the initial condition of the Luenberger observer. The initial condition is generated by a linear interpolation of $\bm{Y}(\bm{x},0)$. Then, at every time step, $\bm{y}(\bm{x},t)$ are converted to $\bm{Y}(\bm{x},t)$, which are used to the observer \eqref{eqn:observer}. Note that the observer system starts with an initial condition, which is far apart from that of the reference system, and at each time step only noisy low-resolution information is supplied to the observer. Yet, as shown in the middle and bottom rows of Fig. \ref{fig:snapshot}, the reconstructed velocity field, $z_1(\bm{x},t)$, is well synchronized with the ground truth ($u_1(\bm{x},t)$).

\begin{figure*}[t]
    \centering
    \includegraphics[width=.32\textwidth]{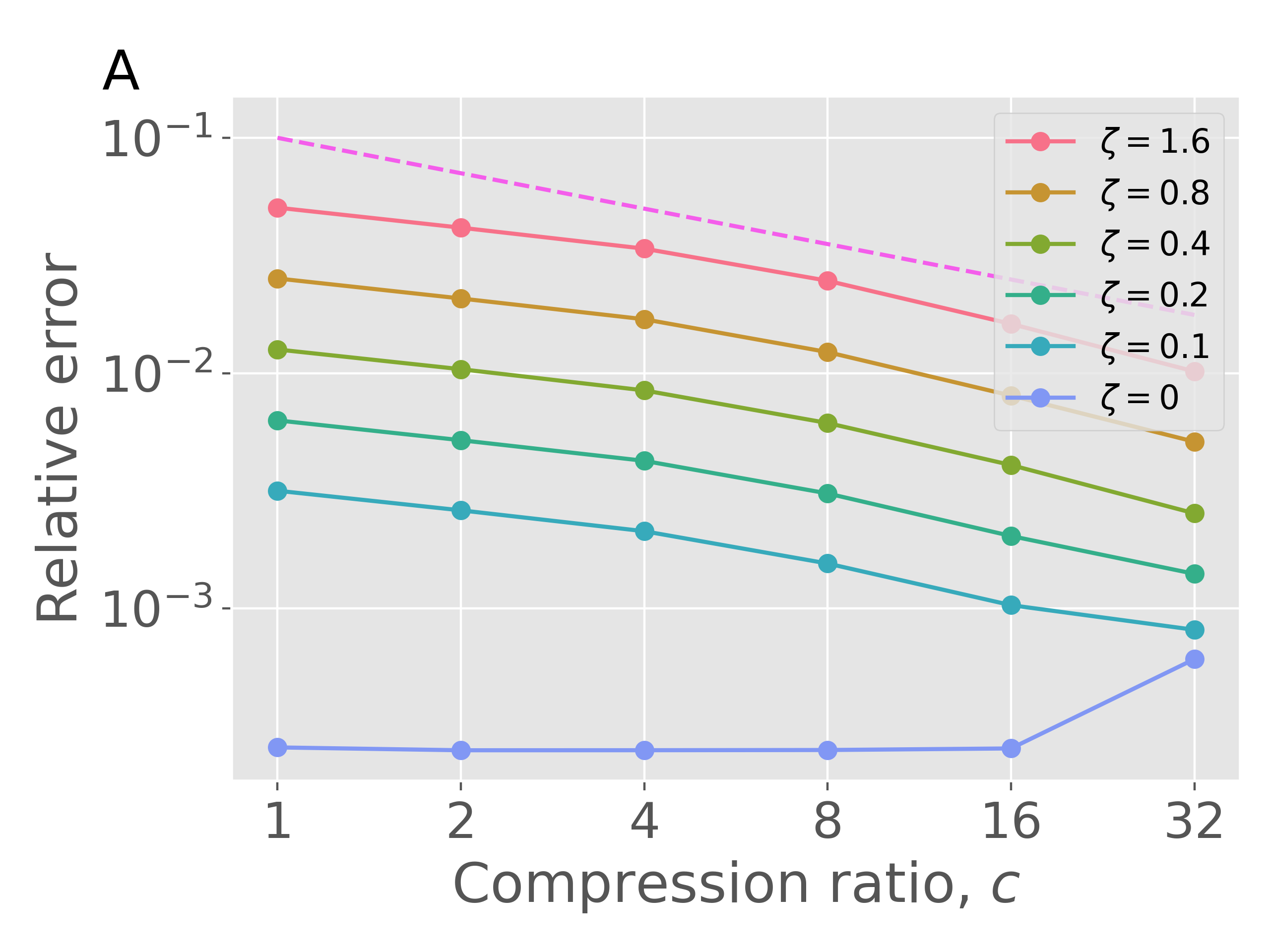}
    \includegraphics[width=.32\textwidth]{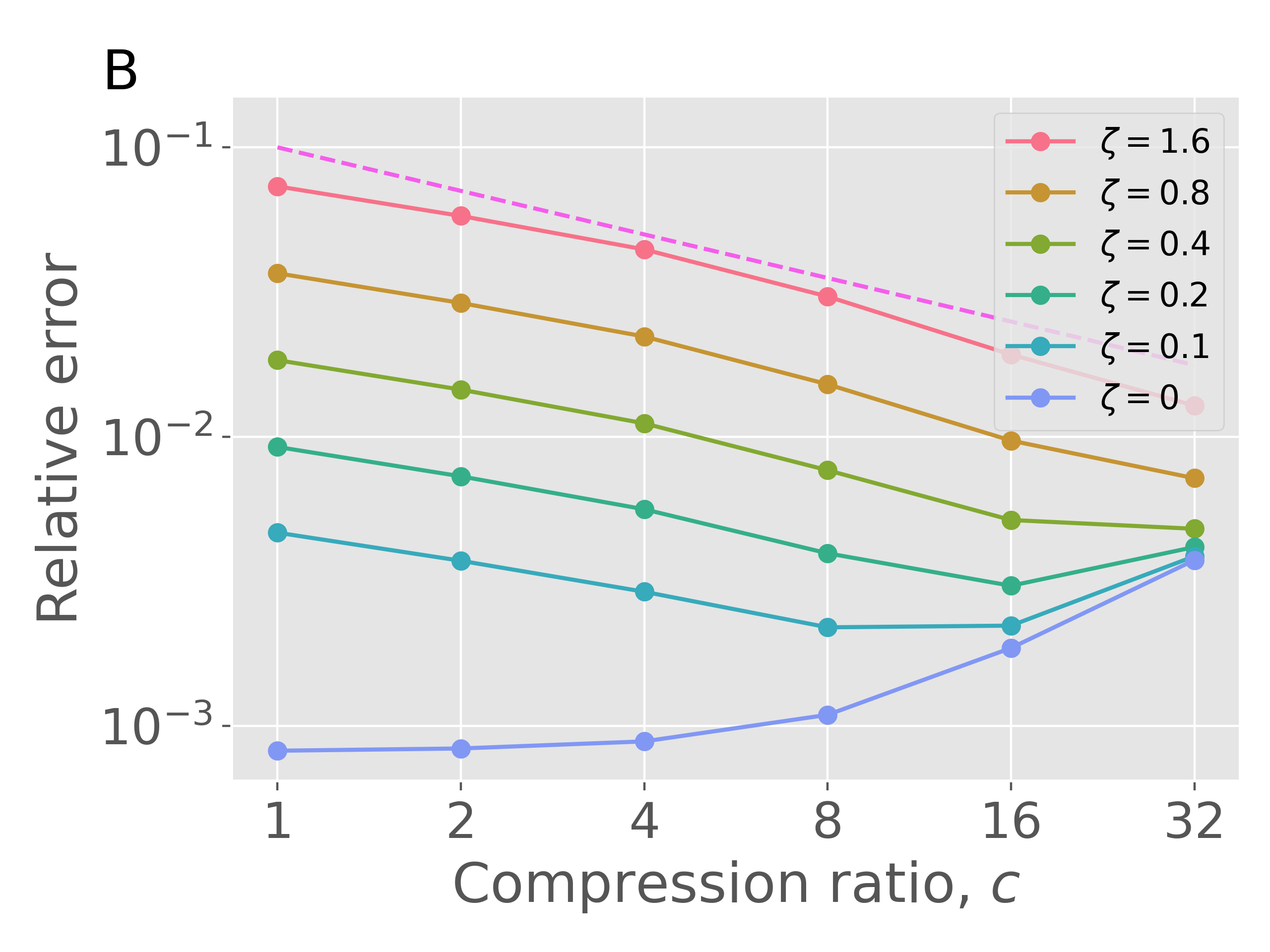}
    \includegraphics[width=.32\textwidth]{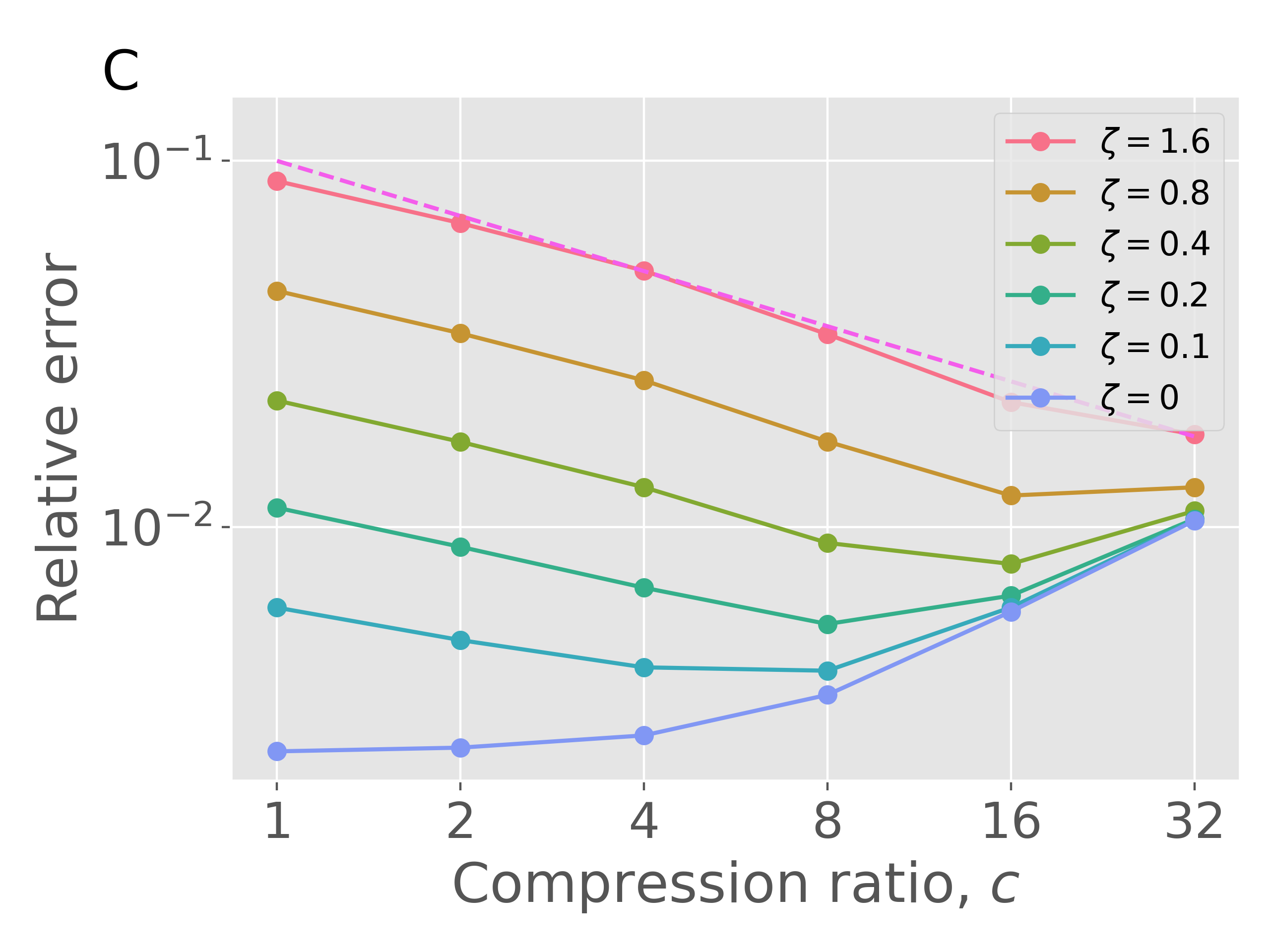}
    \caption{Changes of the relative error with respect to $c$ for $\nu=0.009$ (A), $0.003$ (B), and $0.0015$ (C). $c$ is related to the subdomain for the spatial averaging $|\Omega_i| = (c \delta_x)^2$. $L$ is fixed ($L=10$). The dashed line indicates $\sim c^{-0.5}$.  }
    \label{fig:c_error}
\end{figure*}


From the definitions of $\Upsilon$ and the noise magnitude in (\ref{eq:noise_var}--\ref{eq:tke}), we can define a relative error, $r(t)$, which scales as
\begin{equation} \label{eq:re-scaling}
r(t) = \frac{\|\bm{e}(\bm{x},t) \|_{L^2}}{\|\bm{u} (\bm{x},t) \|_{L^2}}
\end{equation}
Then, the theoretical analysis predicts $r(t) \le \frac{\zeta}{\delta c}$. 

To investigate the dynamics of $\bm{e}$, we first show the temporal changes of $r(t)$ in Fig.~\ref{fig:time_error}. In Fig. \ref{fig:time_error} A, $r(t)$ for $\nu=0.006$, $\zeta=1.6$, and $L=5$ are shown. Here, $c=1$ indicates using the noisy high-resolution data directly to compute $\bm{z}(\bm{x},t)$. It is shown that, for $c \le 16$, indeed $r(t)$ exhibits an exponential decay. For $c=32$, the exponential decay of $r(t)$ is observed after an initial transient behavior for $t<3$. On the other hand, when the size of $\Omega_i$ becomes too large, e.g., $c=64$, the state estimation diverges from the ground truth. Note that when $\bm{u}$ and $\bm{z}$ are totally uncorrelated, $r(t) \simeq 2$. Hence, the result implies that, even for $c=64$, $\bm{z}$ still follows $\bm{u}$, but $\bm{Y}$ does not have enough information to closely reconstruct small-scale eddies of $\bm{u}$. Consistent with the theory, it is also observed that $r(t)$ monotonically decreases with the increase in the size of $\Omega_i$ for $c \le 32$.

Fig. \ref{fig:time_error} B shows $r(t)$ for a lower viscosity, $\nu=0.003$. It is observed that $r(t)$ still shows a relaxation dynamics toward $\Upsilon$. As $\nu$ becomes smaller, or as the Reynolds number increases, small-scale eddies become more energetic, which makes it more challenging to reconstruct $\bm{u}$. Unlike Fig. \ref{fig:time_error} A, $r(t)$ shows a monotonic decrease only up to $c=16$. For $c=32$, $r(t)$ converges much more slowly. Then, $r(t)$ loses such convergence for $c > 32$.

Now, we investigate the effects of $L$ on the estimation error. For this we fix $\nu=0.0015$, $\zeta=1.6$, and $c=32$, and plot estimation error for different values of $L$ in Fig. \ref{fig:time_error} C. It is shown that initially increasing $L$ makes $r(t)$ converge faster. But, for $L \ge 10$, the initial decrease of $r(t)$, for $t < 5$, essentially collapses onto one curve. It is also shown that when $L > 10$ $r(t)$ becomes an increasing function of $L$. This result indicates that there is an optimal range of $L$ to reach the minimum $r(t)$.

Fig. \ref{fig:c_error} shows the changes in the convergence zone in terms of $c$. The experimental convergence zone is computed by a time-average of $r(t)$ after period $t_0$ when $r(t)$ becomes stationary:
\begin{equation}
    E(c)= \frac{1}{2}\int_{t_0}^{t_0+2}r(t)dt.
\end{equation}
As predicted in the theoretical analysis, in general, $E(c)$ is a decreasing function of $c$, clearly indicating that our SR algorithm indeed improves the reconstruction better. However, it is observed that $E(c)$ behaves roughly $\sim c^{-0.5}$, while the theoretical results in \eqref{eqn:conv_zone} predict a faster decay, $\sim c^{-1}$. As shown in Fig. \ref{fig:time_error}, there is an optimal value of $L$, which is a function of $c$, $\zeta$, and $\nu$. It is possible that $E(c)$ exhibits a faster decay with respect to $c$, if the optimal $L$ is used, instead of fixing $L=10$. However, in practice, it is challenging to find the optimal $L$ for each set of simulation parameters.

In Fig. \ref{fig:c_error} B and C, it is shown that for $\nu \le 0.003$, $E(c)$ decreases with respect to $c$ at first and then starts to increase. It is interesting to observe in Fig. \ref{fig:c_error} C that $E(c)$ starts to increase first from the smaller noise, e.g., $\zeta = 0.1$, and $E(c)$ of larger $\zeta$ eventually collapse onto $E(c)$ of $\zeta=0.1$ as $c$ increases. The increase of $c$ indicates lowering the resolution of $\bm{Y}$.In the continuous data assimilation of turbulent flows with noiseless data \cite{Lalescu13,Inubushi23}, it was found that the critical length below which turbulent eddies can be reconstructed is approximately 20 Kolmogorov lengthscale. Similarly, this result suggests the existence of a critical lengthscale required for the chaos synchronization. Mathematically, increasing $c$ over a threshold makes the SR method violate the observer's convergence condition (see Theorem 1 in Supplementary Material). Physically, the results indicate that, as the lengthscale of the averaging becomes larger, eddies bigger than a critical lengthscale are filtered out, which makes the observer deviates from the reference system. Or, by reducing the viscosity (increasing the Reynolds number), the size of the smallest eddies becomes smaller, which requires the legthscale of the averaging reduced for the chaos synchronization.


In Fig. \ref{fig:c_error}, we also show $E(c)$ of the noiseless observation, i.e., $\zeta = 0$, as the noiseless case provides the lower bound for $E(c)$. For $\zeta = 0$, not surprisingly, $E(c)$ has an opposite trend: $E(c)$ is an increasing function of $c$. Note that, even for $c=1$, the reconstruction error is finite (see Fig. \ref{fig:c_error} A), while the theory for the noiseless observation \cite{Zhuk23} predicts convergence to $\bm{u}$ up to numerical precision. The discrepancy is attributed to the use of different numerical methods for generating data $\bm{Y}$ and solving the observer: $\bm{Y}$ is generated by a highly accurate spectral solver, while the Luenberger observer is computed by a second-order finite element solver. 
We also note that the slower decay of the convergence rate, $\sim c^{-0.5}$, contrary to the theoretical prediction of $\sim c^{-1}$, may be in part due to the numerical errors of the low-order numerical solver. It is also observed that $E(c)$ of the noiseless case eventually grows as $c$ becomes larger. This confirms that when the size of $\Omega_j$ becomes too large compared to the small-scale eddies, the detectability is lost.

Table \ref{tbl:error} lists $E(c)$ for a range of parameters together with $E(c)$ computed from linear and cubic interpolations for the low-resolution data ($c=32$). Note that $c=1$ corresponds to a Newtonian relaxation. 
It is shown that our method makes a much better estimation of $\bm{u}(\bm{x},t)$ compared to those interpolation methods. In particular, $E(c)$ of our method with $c=32$ is less than 5\% of the linear or cubic interpolation. It is also shown that $E(c)$ for $c=32$ is less than 20\% of the Newtonian relaxation ($c=1$), i.e., without the de-noising encoder.

\begin{table}[t] 
    \begin{tabular}{l|cccccc}
\hline\hline
$\nu$ & 0.0015 & 0.003 & 0.0045 & 0.006 & 0.0075 & 0.009 \\
\hline
linear & 0.478 & 0.536 & 0.669 & 0.738 & 0.788 & 0.863 \\
cubic & 0.477 & 0.554 & 0.696 & 0.782 & 0.819 & 0.912\\
$c=1$ & 0.088 & 0.072 & 0.066 & 0.055 & 0.052 & 0.045\\
$c=32$ & 0.018 & 0.013 & 0.012 & 0.010 & 0.010 & 0.009\\
\hline\hline
    \end{tabular}

\caption{Comparison of the estimation errors for the noise level $\zeta = 1.6$; linear and cubic indicate linear and cubic interpolations, respectively. $c=1$ and $c=32$ are from the Luenberger observers with $L=10$.}\label{tbl:error}
\end{table}

In summary, we presented a SR algorithm, which first reduces the noise by applying a linear filtering operator to the data, e.g., averaging high-resolution fluid velocity vector fields over a coarse grid, at the expense of losing small-scale features. Then, this low-resolution but less noisy data is fed into the observer which ``reverses back'' the lost information and reconstructs the ground truth. It is shown that, counter-intuitively, the proposed SR approach makes a better estimation than directly using the high-resolution data. 

In this study, one of the first theoretical results on the chaos synchronization in NS flows on corrupted observations is presented. Our theoretical analysis demonstrates that, unlike the noise-free cases \cite{Lalescu13,Zhuk23}, the observer converges into a $L^2(\Omega)$-ball containing the ground truth. The radius of the $L^2(\Omega)$-ball, $\Upsilon$, provides an upper bound of the estimation error, i.e., a worst-case estimation error. $\Upsilon$ is globally and exponentially attractive; estimation error $\|\bm{u}-\bm{z}\|_{L^2}$ decays exponentially fast in time until $\|\bm{u}-\bm{z}\|_{L^2}=\Upsilon$ and it stays under $\Upsilon$ from there on. It is also found theoretically that $\Upsilon$ is a decreasing function of the length scale of the spatial average, $h=c\delta_x$. In other words, the deviation of the observer from the ground truth decreases, as the resolution of the input data is further lowered until a critical lengthscale is reached. The critical lengthscale is related with the viscosity, i.e., the Reynolds number, and the coupling strength, $L$.

Our numerical experiments indeed demonstrate that the time evolution of $\|\bm{e}\|_{L^2}$ exhibits an exponential decay towards $\Upsilon$, and $\Upsilon$ is, in general, a decreasing function of $c$. 
We note, however, that the predicted decrease of $\sim c^{-1}$ of $\| \bm{e} \|_{L^2}$ was not observed. Instead, a slower decay of $\sim c^{-0.5}$ is observed in the numerical experiments. This may be due to the use of different solvers. While the data is generated by using a spectral solver, the Luenberger observer is solved by using a second-order finite element solver, which introduces additional errors, not considered in the theoretical analysis.

The present SR model can be thought of as a data assimilation method, which is used to estimate the flow state from observations \cite{Leoni20,Zaki21}. {\color{black}While many of the data assimilation models require an observer to observe ``spectrally filtered'' velocity field \cite{Lalescu13,Inubushi23,Olson03}, our SR model only requires a spatial averaging, which dramatically reduces the computational complexity and makes it easier to apply for real measurement data. It is worthwhile to note that, as the SR method relies only on ``local'' information, it can be used for the flow estimation in a subdomain.   
Moreover, we would like to emphasize that our theoretical analysis can be extended to another linear operator $\mathcal{C}$ that satisfies the detectability condition (Eq. (9) in Supplementary Material), not just for the spatial-average operator considered in this study. Generalization of the proposed SR method for a wider range of data assimilation problems, such as three-dimensional and/or inhomogeneous NS flows, reconstruction in a subdomain, and different types of denoising operators, is a subject of future research.}

\begin{acknowledgments}
This work was supported in part by the Hartree National Centre for Digital Innovation, a collaboration between STFC and IBM. We thank the anonymous reviewers for their insightful discussions.
\end{acknowledgments}

\bibliography{our_bib}

\end{document}



\title{Reducing data resolution for better super-resolution:\\ {Reconstructing} turbulent flows from noisy observation\\ \textit{Supplementary Material}}

\author{Kyongmin Yeo}
\email[Author to whom all correspondence should be addressed: ]{kyeo@us.ibm.com}
\affiliation{%
 IBM T.J. Watson Research Center, NY, USA
}%

\author{Ma{\l}gorzata J. Zimo\'{n}}
\affiliation{IBM Research - Europe, UK
}
\thanks{M.J.Z contributed equally to this work with K.Y.}
\altaffiliation[Also at ]{Department of Mathematics, University of Manchester, UK}

\author{Mykhaylo Zayats}
\author{Sergiy Zhuk}
\affiliation{
 IBM Research - Europe, Ireland
}

\date{\today}

\maketitle
 

\section*{Problem Setup}
In this section, we reiterate the problem setup briefly described in the letter to provide a robust mathematical description of the problem.

\textbf{Navier-Stokes equations.} We consider data in the form of noisy measurements of two-dimensional turbulent fluid velocity vector fields. Let the ground truth velocity, $\bm{u}(\bm{x},t)$ be the unique solution of the Navier-Stokes equations (NSE) in two spatial dimensions: 
\begin{align}
&\frac{\partial \bm{u}}{\partial t} = -\frac{1}{\rho}\bm{\nabla}p -\bm{u}\cdot \bm{\nabla}\bm{u} + \nu\nabla^2\bm{u}+\bm{f}, \label{eqn:NS} \\
&\bm{\nabla}\cdot\bm{u} = 0, \label{eqn:div_free}
\end{align}
where $\bm{u}(\bm{x},t)$ is the velocity field, $\bm{f}(\bm{x},t)$ is a known exogenous forcing, $p$ is the pressure and $\nu$ is the kinematic viscosity. The variables ($\bm{u}, \bm{f}, p$) are defined on $(\bm{x},t) \in \Omega \times (0,T]$, in which $\Omega$ is a rectangular domain, $\Omega = (-\ell_1/2,\ell_1/2) \times(-\ell_2/2,\ell_2/2)$, $T$ is the final time and the initial velocity $\bm{u}(\bm{x},0)=\bm{u}_0$ is unknown. To simplify the presentation, we assume periodic boundary conditions: $u_i(\ell_1/2,x_2,t) = u_i(-\ell_1/2,x_2,t)$, $u_i(x_1,\ell_2/2,t) = u_i(x_1,-\ell_2/2,t)$ for $i=1,2$. 

For incompressible fluids, the density of the fluid, $\rho$ is a constant and $p$ is not the thermodynamic pressure, but a Lagrange multiplier to make $\bm{u}$ satisfy the mass conservation law, given by \eqref{eqn:div_free}. The Reynolds number of the flow is defined as
\[
Re = \frac{\mathcal{U}\mathcal{L}}{\nu},
\]
in which $\mathcal{U}$ and $\mathcal{L}$ denote the characteristic velocity and length scales, respectively. The Reynolds number denotes the relative importance between the inertial acceleration and the viscous dissipation. For a small $Re$, the viscous dissipation becomes dominant, which effectively suppresses small-scale perturbations, resulting in laminar flows. On the other hand, for $Re \gg 1$, the flow becomes turbulent, which is characterized, e.g., by the presence of small-scale vortices.

\textbf{Low-dimensional projection.} Assume that $\Omega$ is partitioned into $N$ disjoint rectangles: 
\begin{equation}\label{eq:Omega}
    \Omega = \bigcup_{j=1}^N \Omega_j,~ \Omega_j = (a_j,a_j+h^j_1)\times(b_j,b_j+h^j_2)\subset\Omega\,.
\end{equation}
Let $|\Omega_j|=h^j_1h^j_2$ (the area of $\Omega_j$), and define an indicator function for $\Omega_j$, $\xi_j(\bm{x})=1$ if $\bm{x}\in\Omega_j$ and $\xi_j(\bm{x})=0$ otherwise. This partition defines the sampling operator $\oC$ which relates $\bm{u}$ to a low-resolution data $\bm{Y}$ by averaging $\bm{u}$ over $\Omega_j$: 
\begin{align} 
\bm{Y} &= \oC(\bm{u}+\bm{\eta}),\label{eq:Y}\\
(\mathcal{C}\bm{u})(\bm{x},t) &= \sum_{j=1}^N \left(\frac{1}{|\Omega_j|}
    \int_{\Omega_j}\bm{u}(\bm{x},t)~ d\bm{x}\right) \xi_j(\bm{x}).\label{eq:Coverages}
\end{align}
$\bm{\eta}$ is a white noise, which models the observation noise. It is assumed that $\bm{\eta}$ is bounded: 
\begin{equation} \label{eq:noise_bound}
    \sum_{j=1}^N R_{k,j}(t) \left( \frac{1}{|\Omega_j|} \int_{\Omega_j} \eta_k(\bm{x},t) d\bm{x} \right)^2 \le 1,
\end{equation}
for $t \in(0,T)$, $R_{k,j}(t)>R>0$, and $k=1,2$. The weights $R_{k,j}$ determine the amount of noise in each $\Omega_j$: the larger $R_{k,j}$, the smaller (on average) the noise in $\Omega_j$.  


\textbf{Luenberger observer.}
Let $\bm{z}(\bm{x},t)$ denote the unique solution of the Luenberger observer \cite{Zhuk23}: 
\begin{align}
&\frac{\partial \bm{z}}{\partial t} = -\frac{1}{\rho}\bm{\nabla}p -\bm{z}\cdot \bm{\nabla}\bm{z} + \nu\nabla^2\bm{z}+\bm{f}+L(\bm{Y}-\mathcal{C}\bm{z}), \label{eqn:observer} \\
&\bm{\nabla}\cdot\bm{z} = 0. \label{eqn:div_freez}
\end{align}
It was demonstrated in~\cite{Zhuk23} that, when the data is noise-free, i.e, $\bm{\eta} = \bm{0}$, then $\bm{z}$ converges to the ground truth $\bm{u}$ over time in the norm of Sobolev space $H^1(\Omega)$ independently of the initial velocity $\bm{u}(\bm{x},0)$. This is true provided that the data generation process is detectable: namely the area of the largest $\Omega_j$, $h^2=\max_j(\max\{h_1^j,h_2^j\})^2$ verifies a certain inequality depending on $L$ and $\nu$, and at the same time it satisfies 
\begin{equation}\label{eq:detectC}
\exists~h>0, C_\Omega>0:\quad \|\bm{u} - \oC\bm{u}\|_{\vsLt}^2\le h^2 C_\Omega  \|\nabla\bm{u}\|_{\vsLt}^2,
\end{equation}
which is always the case for $\oC$ defined by~\eqref{eq:Coverages}, thanks to the Poincar\'{e} inequality. More generally, the data generation process is detectable for a broad class of linear operators $\oC$ verifying~\eqref{eq:detectC}, not just for averaging operators like~\eqref{eq:Coverages}~\cite{Zhuk23}. 


\section*{Proof of Convergence of the observer}\label{sec:proof}
In this section, we show that in the case of noisy data, the observer converges into a ball containing the ground truth and provides an upper bound for its radius. Next, we offer an intuitive explanation of how increasing the size of rectangles $\Omega_j$ reduces this upper bound. 

\subsection{Preliminary}
Let $\R^n$ denote Euclidean space of dimension $n$ with inner product $\vu\cdot\bm v = \sum_{i=1}^n u_i v_i$, $\R^n_+$ -- non-negative orthant of $\R^n$, $\R^1_+=\R_+$, and for $\vk,\vL\in\R^n$ set $\frac{\vk}{\vL}=(\frac{k_1}{\ell_1}\dots \frac{k_n}{\ell_n})^\top$. Let $\sLCO(H)$ denote the space of all closed linear operators $\oC$ acting in a Hilbert space $H$ with domain $\dom(\oC)\subset H$. The following functional spaces are standard in NSE's theory (~\cite[p.45-p.48]{foias2001}):\begin{itemize}
\item $L_p^2(\Omega)$ -- space of $\Omega$-periodic functions $u:\Omega\subset\R^2\to\R$ with period $\Omega=(-\frac{\ell_1}2, \frac{\ell_1}2)\times (-\frac{\ell_2}2, \frac{\ell_2}2)$ for some $\ell_{1,2}>0$
  and inner product $(w,v)=\int_\Omega wv dx_1 dx_2$
        \item $L_p^2(\Omega)^2$ -- space of $\Omega$-periodic vector-functions $\vu=\left[\begin{smallmatrix}
            u_1 u_2
          \end{smallmatrix}\right]$ with inner product $(\vu,\bm\phi) = (u_1,\phi_1)+(u_2,\phi_2)$ and norm $\|\vu\|^2_{\vsLt}=\|u_1\|^2_{\vsLt}+\|u_1\|^2_{\vsLt}$
	\item $H_p^1(\Omega)=\{u\in L_p^2(\Omega): \|\nabla u\|_{\R^2}\in L_p^2(\Omega)\}$,\\ $H_p^1(\Omega)^2=\{\vu=
          \left[\begin{smallmatrix}
            u_1 u_2
          \end{smallmatrix}\right]\in L_p^2(\Omega)^2: u_{1,2}\in H_p^1(\Omega)\}$ with norm $\|\vu\|^2_{H^1}=\|\vu\|_{\vsLt}^2+\|\nabla\vu\|^2_{\vsLt}$ where $\|\nabla\vu\|^2_{\vsLt}=\int_\Omega \|\nabla \vu\|_{\R^2}^2dx_1dx_2$, and $\|\nabla \vu\|_{\R^2}^2=\|\nabla u_1\|_{\R^2}^2+\|\nabla u_2\|_{\R^2}^2$
	\item $\vsH = \{\bm v \in [L_p^2(\Omega)]^2:\,\nabla\cdot\bm v = 0\}$ -- space of divergence-free vector-functions $\bm v$, $\vsHza = \{\bm v \in \vsH : \quad \int_\Omega \bm v d\bm{x} =0\}$ -- subspace of $\vsH$ of $\bm v$ with zero mean components
        \item $\vsV = \{\bm v\in [H_p^1(\Omega)]^2 :\quad \nabla\cdot\bm v = 0\}$ and $\vsVza = \{\bm v \in \vsV: \quad \int_\Omega \bm v d\bm{x} =0\}$
        \item $L^2(0,T,H)$ -- space of $H$-valued functions $t\mapsto u(t)\in H$ with finite norm $\|u\|^2_{L^2(0,T,H)}=\int_0^T \|u(t)\|_H^2 dt$ for $T\in(0,+\infty)$, e.g. $L^2(0,T,\vsHza)$ -- space of $\bm v(x_1,x_2,t)$ such that $\int_0^T \int_\Omega \|\bm v(x_1,x_2,t)\|_{\R^2}^2dx_1dx_2 <+\infty$ and $\bm v(\cdot,t)$ has zero divergence and zero mean for almost all $t\in (0,T)$
        \item $L^\infty(0,T,H)$ -- space of $H$-valued functions $t\mapsto u(t)\in H$ such that $\|u(t)\|_H \le C<+\infty$ for some $C>0$ and almost all $t\in (0,T)$, $T\in(0,+\infty)$ with finite norm $\|u\|^2_{L^\infty} = \|u\|^2_{L^\infty(0,T,H)}=\min_{C}\{C>0: \|u(t)\|_H \le C\}$
        \end{itemize}

\subsubsection{Bounds for $L^\infty$-norms of periodic vector-functions}
\label{sec:upper-bounds}

The following lemma from~\cite{Zhuk23} is used below to prove the convergence theorem given below. 

\begin{lemma}\label{sec:upper-bounds-linfty}
If $\vu\in H_p^2(\Omega)^2$ and\footnote{This condition is necessary: lemma does not hold for $u\equiv \text{const}$ and small enough $\ell_{1,2}$.} $\int_\Omega \vu dx_1dx_2=0$ then for any $\gamma>0$ it holds:
\begin{equation}
  \label{eq:AgmoN}
  \begin{split}
  \|\vu\|_{L^\infty}&\le \frac{\log^{\frac12}\left(1+\frac{4\pi^2 \gamma^2}{\ll_1\ll_2}\right)\|\vu\|_{H^1}}{\sqrt{2\pi}} + \frac{\|\vL\|_{\R^2} \|\Delta \vu\|_{L^2}}{\gamma\sqrt{32\pi^3}}.
\end{split}
\end{equation}
For $\gamma= \|\vu\|_{H^1}^{-\frac12}\|\Delta \vu\|_{L^2}^{\frac12}$, \eqref{eq:AgmoN} gives 2D Agmon's inequality~\cite[p.100]{foias2001}:
\begin{equation}
  \label{eq:Agmon}
  \begin{split}
  \|\vu\|_{L^\infty}&\le \left(\sqrt{\frac{2\pi}{\ll_1\ll_2}}+\frac{\|\vL\|_{\R^2}}{\sqrt{32\pi^3}}\right)\|\vu\|_{H^1}^{\frac12}\|\Delta \vu\|_{L^2}^{\frac12}.
\end{split}
\end{equation}
For $\gamma= \|\vu\|_{H^1}^{-1}\|\Delta \vu\|_{L^2}$, \eqref{eq:AgmoN} gives 2D Brezis inequality:
\begin{equation}
  \label{eq:Brezis}
  \|\vu\|_{L^\infty}\le \left(\frac{\|\vL\|_{\R^2}}{\sqrt{32\pi^3}}+\frac{\log^{\frac12}(1+\frac{4\pi^2 \|\Delta \vu\|_{L^2}^2}{\ll_1\ll_2 \|\vu\|_{H^1}^2})}{\sqrt{2\pi}}\right)\|\vu\|_{H^1}.
\end{equation}
\end{lemma}

\subsection{Navier-Stokes equation: weak  formulation and well-posedness in 2D}

Let us transform~\eqref{eqn:NS} into Leray's weak formulation: to this end we eliminate pressure $p$ by multiplying~\eqref{eqn:NS} by a test function $\bm\phi\in\vsVza$, and integrate by parts in $\Omega$ to obtain Leray's weak formulation of NSE in 2D:
	\begin{equation}
	\label{eq:NSE-var}
	\dfrac{d}{dt}(\vu,\bm \phi) + b(\vu,\vu,\bm\phi) + \nu((\bm u,\bm \phi)) = (\bm f,\bm \phi), \quad\forall\bm\phi\in \vsVza
      \end{equation}
      with initial condition $(\vu(0),\bm\phi)=(\vu_0,\bm\phi)$. Here
\begin{align*}
 b(\bm u,\bm w,\bm\phi) = (\bm u\cdot \nabla w_1,\phi_1) + (\bm u\cdot \nabla w_2,\phi_2),
                             ((\bm u,\bm \phi)) = (\nabla u_1,\nabla \phi_1)+(\nabla u_2,\nabla \phi_2).
\end{align*}
In what follows we will be using some properties of the trilinear form $b$ and Stokes operator  $\vu\mapsto A\vu=-P_l\Delta \vu$, a self-adjoint positive operator with compact inverse, which coincides with $\Delta \vu$ for periodic BC (see~\cite[p.52]{foias2001}): for $\vu\in \dom(A)$ and $\vphi\in \vsVza$
\begin{align}
  (A\bm u, \vphi) &= ((\bm u, \vphi)),\,(A\bm u,\bm u) = ((\bm u,\bm u))\ge \lambda_1(\vu,\vu)  \label{eq:Poincare},\\
  (A\bm u, A\bm u)  &=(A(A^\frac12)\bm u,(A^\frac12)\bm u)\ge \lambda_1 (A \bm u, \bm u)\label{eq:gradA-bnd},\\
  \lambda_1&=4\pi^2/\max\{\ell_1,\ell_2\}^2,\\
 b(\bm u,\bm v,\bm \phi) &= - b(\bm u,\bm \phi,\bm v),\\
 b(\bm u,\bm v,\bm v) &=0,\quad  b(\bm v,\bm v,A\bm v) = 0.\label{eq:b:ort}
\end{align}

Next Lemma collects results from~\cite[p.58, Th.7.4, p.99, f.(A.42), p.102, f.(A.66)-(A.67)]{foias2001} on existence, uniqueness, regularity and input-to-state stability of NSE's weak (strong) solution $\vu$, and bounds for $A\vu$. Classical smoothness of $\vu$ requires further constraining of $\vf$, $\vu_0$~\cite[p.59]{foias2001}. 
\begin{lemma}\label{l:2}
	Let $\vu_0\in \vsHza$ and $\vf\in L^2(0,\bar T,\vsHza)$. Then, on $[0,\bar T]$ there exist the unique weak solution $\vu\in C(0,\bar T,\vsHza)$ of NSE~\eqref{eq:NSE-var}, and the components of $\vu=\begin{smallmatrix}[u_1 & u_2]\end{smallmatrix}$ verify: $u_i,(u_i)_{x_1,x_2}\in L^2(\Omega\times(0,\bar T))$. If $\vu_0\in\vsVza$ then the weak solution coincides with the strong solution of~\eqref{eq:NSE-var}, and \[
	\dfrac{du_i}{dt}, (u_i)_{x_1},(u_i)_{x_2}, (u_i)_{x_1 x_2}\in L^2(\Omega\times(0,\bar T)),
      \] i.e., $\vu\in C(0,\bar T,\vsVza) \cap L^2(0,\bar T,\mathscr{D}(A))$. If in addition $\bm f\in \vsLiHza$ then $(\,\|\bm f\|_{L^\infty} = \|\bm f\|_{\vsLiHza}$ for short$)$:
	\begin{align}
          &\|\bm u(\cdot,t)\|^2_{\vsLt} \le \frac{\|\bm f\|^2_{L^\infty}}{(\nu\lambda_1)^2} + e^{(-\lambda_1 \nu)(t-s)}\|\bm u(\cdot,s)\|^2_{\vsLt}	\label{eq:L2norm-decay}\\
	&\|\nabla \bm u(\cdot,t)\|^2_{\vsLt} \le \frac{\|\bm f\|^2_{L^\infty}}{\nu^2\lambda_1} + e^{(-\lambda_1 \nu)(t-s)}\|\nabla \bm u(\cdot,s)\|^2_{\vsLt}          	\label{eq:grad-norm-decay}\\
          &\frac1{\bar T}\int_t^{\bar T+t}\|A\vu\|^2_{\vsLt}ds  \le \theta_{t,\bar T}:=\frac{2\|\vf\|^2_{L^\infty}}{\bar T\nu^3\lambda_1} + \frac1{\bar T}\int_t^{\bar T+t}\frac{\|\vf(s)\|^2_{\vsLt}}{\nu^2}ds + \frac{2 e^{(-\lambda_1 \nu)t}\|\nabla\bm u_0\|^2_{\vsLt}}{\bar T\nu}\label{eq:Au-ThetatT}
        \end{align}
\end{lemma}

\subsection{Convergence Theorem}
Recall definitions of $\bm{Y}$, $\oC$ and the partition of the domain $\Omega$ from~\eqref{eq:Omega}. We first provide a formulation of Convergence Theorem using the weak form of the NSE and observer: 

\begin{theorem*}[Observer convergence]
Let $\vu$ solve NSE with $\vf=\vg+\bm d$:
\[
  \dfrac{d}{dt}(\bm u,\bm \phi) + b(\vu,\vu,\bm\phi) + \nu((\vu,\bm \phi)) = (\vg+\bm d,\bm \phi),\quad \vu(0)\in\vsVza
  \]
  and $\vz$ solve observer \[
\dfrac{d}{dt}(\bm z,\bm \phi) + b(\bm z,\bm z,\bm\phi) + \nu((\bm z,\bm \phi)) = (\bm F,\bm \phi),\quad \vz(0)=0
  \]
with \[
      \vF = \vg+L\bm{Y} - L \oC \vz = \vg + L \oC \ve + L \oC \bm \psi\,.
    \]
    Take $\alpha,\delta\in(0,1)$ and define $\Sigma_{h,R} = 2h^2\max_k \max_j R^{-1}_{k,j}$
\begin{align*}
       C_u(t,T_1,\Gamma)&:=2\frac{\|\bm{\ell}\|^2_{\R_2}}{32\pi^2\Gamma^2}+2\log\left(1+\frac{4\pi^2 \theta_{t,T_1} \Gamma^2}{\ll_1\ll_2}\right)
       \times \left(\frac{(1+\lambda_1)\|\bm{f}\|^2_{L^\infty}}{(\nu\lambda_1)^2}+  e^{(-\lambda_1 \nu)t }\|\bm{u}_0\|^2_{H^1_p(\Omega)} \right),\\
       \theta_{t,\bar T} &= 2\frac{\|\bm{f}\|^2_{L^\infty}}{\bar T\nu^3\lambda_1} 
       +\frac1{\bar T}\int_t^{\bar T+t}\frac{\|\bm{f}(s)\|^2_{\vsLt}}{\nu^2}ds
      + \frac{2 e^{(-\lambda_1 \nu)t}\|\nabla\bm{u}_0\|^2_{\vsLt}}{\bar T\nu},
   \end{align*}
in which $\bm{u}_0 = \bm{u}(\bm{x},0)$.
Then, there exist $h>0$ and $L>0$ (e.g.~\eqref{eq:hL}) such that $\bm{z}$ converges exponentially fast (in time) into a vicinity of $\bm{u}$, namely, there exists time $t^\star>0$ such that 
\begin{equation}\label{eq:zone2}
    \forall t>t^\star:\quad \|\bm{u}(\cdot,t)-\bm{z}(\cdot,t)\|_{L^2(\Omega)} \le \frac{\Sigma^{\frac12}_{h,R}}{\delta}
\end{equation}
and for $t<t^\star$, the estimation error $\|\bm{u}(\cdot,t^\star)-\bm{z}(\cdot,t^\star)\|_{L^2(\Omega)}$ decays exponentially fast towards $\Sigma_{h,R}^{\frac 12}$. Within the zone in \eqref{eq:zone2}, the error may behave non-monotonically, but it will never exceed $\Sigma_{h,R}^{\frac 12}$: in other words, zone \eqref{eq:zone2} is globally exponentially attracting.  
\end{theorem*}

\begin{proof}
Note that for $h^2=\max_j(\max\{h_{x_1}^j,h_{x_2}^j\})^2$ and $C_\Omega=(4\pi^2)^{-1}$ by Poincaré inequality~\eqref{eq:Poincare} we get: 
\begin{equation}\label{eq:detectC2}
\|\ve - \oC\ve\|_{\vsLt}^2\le F(h, C_\Omega)  \|\nabla\ve\|_{\vsLt}^2,\quad  F(h,C_\Omega) =   h^2 C_\Omega
\end{equation}
For simplicity, assume that $h=h_{x_1}^j=h_{x_2}^j$ so that $h^2 = \int_{\Omega_j} d \bm x$. Subtracting the observer equation from NSE one finds that the dynamics of the error $V=(\ve,\ve)$, $\ve=\vu-\vz$ is governed by
    \begin{equation}
      \label{eq:error-Ae1}
      \dfrac{d}{dt}(\bm e,\bm e) + \nu((\bm e,\bm e)) = (\vf-\bm F,\bm e) + b(\bm e,\bm e, \bm u).
      \end{equation}
      Let us transform~\eqref{eq:error-Ae1}: for any $\Lambda_{1,2}>0$ and $q^2=2\|\vf-\vg\|^2_{\vsLt}=2\|\bm d\|^2_{\vsLt}$ %
   \begin{align}
     (\vf-\bm F,\bm e) \le& (\bm d + L(\ve-\oC\ve),\ve) - L(\ve,\ve)-L(\oC \bm \psi, \ve) \nonumber \\
    &+ \Lambda_1(F(h, C_\Omega) \|\nabla\ve\|_{\vsLt}^2 - \|\ve - \oC\ve\|_{\vsLt}^2)
    + \Lambda_2(q^2(t) - \|\bm d\|_{\vsLt}^2), 
   \end{align}
   where  $\ve = (e_1,e_2)^\top$. Note that:
\begin{align*}
  (\oC \bm \psi, \ve) &= \sum_{j=1}^N \left(\frac1{h^2}\int_{\Omega_j} \eta_1 d \bm x \int_{\Omega_j}e_1d\bm x  + \frac1{h^2}\int_{\Omega_j} \eta_2 d \bm x \int_{\Omega_j} e_2d\bm x\right)\\
                       &\le \sum_{j=1}^N \frac{\sqrt{R_{1,j}}}{h^2}\int_{\Omega_j} \eta_1 d \bm x \frac{1}{\sqrt{R_{1,j}}}\int_{\Omega_j} e_1d\bm x  + \frac{\sqrt{R_{2,j}}}{h^2}\int_{\Omega_j} \eta_2 d \bm x \frac{1}{\sqrt{R_{2,j}}}\int_{\Omega_j} e_2d\bm x\\
&\le\sum_{k=1}^2\left(\sum_{j=1}^N \frac{R_{k,j}}{h^4}\left(\int_{\Omega_j} \eta_k d \bm x\right)^2\right)^{\frac12} \left(\sum_{j=1}^N\frac{1}{R_{k,j}}\left(\int_{\Omega_j}e_kd\bm x\right)^2\right)^{\frac12}\\
                                    &\le \sum_{k=1}^2\left(\sum_{j=1}^N \frac{R_{k,j}}{h^4}\left(\int_{\Omega_j} \eta_k d \bm x\right)^2\right)^{\frac12} \left(\sum_{j=1}^N\frac{h^2}{R_{k,j}}\int_{\Omega_j}e^2_kd\bm x\right)^{\frac12}\\
                                    &\le \sqrt{2}h \left(\sum_{k=1}^2\sum_{j=1}^N\frac{1}{R_{k,j}}\|e_k\|^2_{L^2(\Omega_j)}\right)^{\frac12}\\
                                    &\le  \sqrt{2}h\left(\sum_{j=1}^N(R^{-1}_{1,j}\|e_1\|^2_{L^2(\Omega_j)} + R^{-1}_{2,j}\|e_2\|^2_{L^2(\Omega_j)})\right)^{\frac12}\\
                            &\le \sqrt{2}h\max_k \max_j R^{-\frac12}_{k,j} \|\ve\|_{\vsLt}\\
  &\le \Sigma^{\frac 12} \|\ve\|_{\vsLt}, \quad \Sigma_{h,R} = 2h^2\max_k \max_j R^{-1}_{k,j}.
\end{align*}
where to go from 2nd to 3rd line we used Cauchy-Schwartz inequality, and Jensen inequality $\left(\int_{\Omega_j}e_kd\bm x\right)^2\le h^2 \int_{\Omega_j}e^2_kd\bm x$ was used to go from 3rd to 4th. As a result, we get: \[
- L(\ve,\ve)-L(\oC \bm \psi, \ve) \le -L (\ve, \ve) + L \Sigma_{h,R}^{\frac 12} \|\ve\|_{L^2(\Omega)}.
\] Hence, for a $\delta<1 $ and $\delta \|\ve\|_{L^2(\Omega)} > \Sigma_{h,R}^{\frac 12}$ we get: \[
L \Sigma^{\frac 12} \|\ve\|_{L^2(\Omega)} < L \delta \|\ve\|^2_{L^2(\Omega)}
\]
and so \[
- L(\ve,\ve)-L(\oC \bm \psi, \ve) \le -L (1-\delta) \|\ve\|^2_{L^2(\Omega)}, \text{ provided } \delta \|\ve\|_{L^2(\Omega)} > \Sigma_{h,R}^{\frac 12}.
  \]
  Therefore, whenever the error $\ve$ is such that $\delta \|\ve\|_{L^2(\Omega)} > \Sigma_{h,R}^{\frac 12}=\sqrt{2}h\max_k \max_j R^{-1/2}_{k,j}$ we have
  \begin{align}
     (\vf-\bm F,\bm e) \le & (\bm d + L(\ve-\oC\ve),\ve) -L (1-\delta) \|\ve\|^2_{L^2(\Omega)} \\ \nonumber
        &+ \Lambda_1(F(h, C_\Omega) \|\nabla\ve\|_{\vsLt}^2 - \|\ve - \oC\ve\|_{\vsLt}^2)
        + \Lambda_2(q^2(t) - \|\bm d\|_{\vsLt}^2). \label{eq:vgLA}
   \end{align}
Recall definition of $b$: we have for $0\le\alpha\le1$
   \begin{align*}
     -\alpha&\nu((\bm e,\bm e))  + b(\bm e,\bm e, \bm u)
     = \sum_{k=1}^2\int_\Omega (-\nu\alpha (\nabla e_k\cdot \nabla e_k)+ (\bm e\cdot \nabla e_k)u_k d)\bm x\\
                                                             &= - \| \sqrt{\nu\alpha} \nabla e_1 - \sqrt{f_1} u_1\ve\|^2_{L^2(\Omega)} - \| \sqrt{\nu\alpha} \nabla e_2 - \sqrt{f_2} u_2\ve\|^2_{L^2(\Omega)}\\
                                                             &+ f_1 \|u_1\ve\|^2_{L^2(\Omega)} + f_2 \|u_2\ve\|^2_{L^2(\Omega)} \text{, provided }2 \sqrt{f_1 \nu\alpha} = 1 = 2 \sqrt{f_2 \nu\alpha}\\
                                                             &\le \frac1{4\alpha\nu } \int_\Omega \|\vu(\bm x,t)\|_{\R^2}^2(\ve(\bm x,t)\cdot\ve(\bm x,t))d\bm x\\
     &\le \frac{\|\vu(t)\|^2_{L^\infty(\Omega)}}{4\alpha\nu } \|\ve(t)\|^2_{L^2(\Omega)}.
   \end{align*}
Finally, we find:
   \begin{align}
     \dfrac{d}{dt}(\bm e,\bm e) &= -\alpha\nu((\bm e,\bm e)) + b(\bm e,\bm e, \bm u)  - (1-\alpha)\nu ((\bm e,\bm e))  + (\vf-\bm F,\bm e)\\
                                  &\le  -((1-\alpha)\nu-\Lambda_1F(h, C_\Omega)) ((\bm e,\bm e)) + \frac{\|\vu(t)\|^2_{L^\infty(\Omega)}}{4\alpha\nu } \|\ve(t)\|^2_{L^2(\Omega)}\\
                                  &\quad  +L((\ve-\oC\ve),\ve) - \Lambda_1 \|\ve - \oC\ve\|_{\vsLt}^2 + \Lambda_2(q^2(t) - \|\bm d\|_{\vsLt}^2) + (\bm d ,\ve) - f_3\|\ve\|^2_{\vsLt}\\
                                  &\quad +(- L (1-\delta) +f_3)\|\ve\|^2_{L^2(\Omega)}\\
                                  &=-\|\sqrt{\lambda_1}\sqrt{(1-\alpha)\nu -\Lambda_1F(h, C_\Omega)}\ve - \Lambda_1^\frac12(\ve - \oC\ve)\|^2_{\vsLt}-\|\Lambda_2^\frac12 \bm d -f_3^\frac12 \ve\|^2_{\vsLt}\\
                                  &\quad+(- L (1-\delta) + \frac{\|\vu(t)\|^2_{L^\infty(\Omega)}}{4\alpha\nu }  + f_3)\|\ve\|^2_{L^2(\Omega)}+\Lambda_2q^2(t),
   \end{align}
   provided \[
2 \sqrt{\lambda_1}\sqrt{(1-\alpha)\nu -\Lambda_1F(h, C_\Omega)}\Lambda_1^\frac12 = L ,\quad 2 \Lambda_2^\frac12 f_3^\frac12 = 1,
\] or equivalently \[
0 = -4 \lambda_1 (1-\alpha)\nu \Lambda_1 + 4 \lambda_1 F(h, C_\Omega) \Lambda_1^2 + L^2, (1-\alpha)\nu -\Lambda_1F(h, C_\Omega)\ge 0,\quad f_3 = \frac1{4\Lambda_2},
\]
or
\begin{align*}
  \Lambda_1^{\pm} & = \frac{4 \lambda_1 (1-\alpha)\nu \pm \sqrt{16 \lambda_1^2 (1-\alpha)^2\nu^2-16 \lambda_1 F(h, C_\Omega) L^2 }}{8 \lambda_1 F(h, C_\Omega)} \\
  &=  \frac{(1-\alpha)\nu \pm \sqrt{(1-\alpha)^2\nu^2 - F(h, C_\Omega) L^2\lambda_1^{-1}}}{2 F(h, C_\Omega)}.
\end{align*}
Note that \[
\Lambda_1^{+} = \frac{(1-\alpha)\nu}{2 F(h, C_\Omega)} + \frac{\sqrt{(1-\alpha)^2\nu^2 - F(h, C_\Omega) L^2\lambda_1^{-1}}}{2 F(h, C_\Omega)}\le \frac{(1-\alpha)\nu}{ F(h, C_\Omega)}
\] with equality attained at $L^2=0$ so that $(1-\alpha)\nu -\Lambda_1F(h, C_\Omega)\ge 0$ holds true. Of course, $\Lambda_1^{+}$ makes sense only if \[
(1-\alpha)^2\nu^2 - F(h, C_\Omega) L^2\lambda_1^{-1}\ge 0 \Leftrightarrow h L \le \frac{2\pi}{\max\{\ell_1,\ell_2\}}(1-\alpha)\nu 2\pi.
\]
Hence, \[
  \dfrac{d}{dt}(\bm e,\bm e) \le (- L (1-\delta) + \frac{\|\vu(t)\|^2_{L^\infty(\Omega)}}{4\alpha\nu }  + f_3)\|\ve\|^2_{L^2(\Omega)}+\Lambda_2q^2(t),
  \]
  provided \[
    h L \le \frac{4\pi^2(1-\alpha)\nu}{\max\{\ell_1,\ell_2\}}.
  \]
  Consider the case of known input: $f_3=\Lambda_2=0$. Then \[
    \dfrac{d}{dt}(\bm e,\bm e) \le (\frac{\|\vu(t)\|^2_{L^\infty(\Omega)}}{4\alpha\nu}  - L (1-\delta))\|\ve\|^2_{L^2(\Omega)}.
  \]
  And $(\bm e,\bm e)$ decays provided \[
\frac1{T_1}\int_t^{T_1+t} \frac{\|\vu(s)\|^2_{L^\infty(\Omega)}}{4\alpha\nu} ds <  L (1-\delta).
  \]
   Let us also note that by Lemma~\ref{sec:upper-bounds-linfty}: 
   \[
      \|u(t)\|_{L^\infty(\Omega)} \le \log^{\frac12}\left(1+\frac{4\pi^2 \|A \vu\|_{\vsLt}^2 \Gamma^2}{\ll_1\ll_2}\right)\|\vu\|_{H^1}
      + \frac{\|\bm\ell\|_{\R_2}}{\sqrt{32\pi^2}\Gamma}\label{eq:bAgmon1}
      \]
      for any $\Gamma>0$, so that
        \begin{align*}
      &\frac 1{T_1}\int_{t}^{t+T_1}\|u(\tau)\|^2_{L^\infty(\Omega)} d\tau \\
      &\hspace{2em} \le 2\frac 1{T_1}\int_{t}^{t+T_1} \log\left(1+\frac{4\pi^2 \|A\vu(\tau)\|^2_{L^2} \Gamma^2}{\ll_1\ll_2}\right) \|\vu(\tau)\|^2_{H^1} d\tau     + 2\frac{\|\bm\ell\|^2_{\R_2}}{32\pi^2\Gamma^2}          .  
        \end{align*}
It follows from Lemma~\ref{l:2} that
\begin{align*}
    \|\vu(t)\|^2_{H^1} &\le \frac{\|\bm f\|^2_{L^\infty}}{(\nu\lambda_1)^2} +  e^{(-\lambda_1 \nu)t }\|\bm u_0\|^2_{\vsLt}+\frac{\|\bm f\|^2_{L^\infty}}{\nu^2\lambda_1} + e^{(-\lambda_1 \nu)t}\|\nabla \bm u_0\|^2_{\vsLt}\\
    &=\frac{(1+\lambda_1)\|\bm f\|^2_{L^\infty}}{(\nu\lambda_1)^2} +  e^{(-\lambda_1 \nu)t }\|\bm u_0\|^2_{H^1_p(\Omega)}.
\end{align*}
Hence,\[
    \frac 1{T_1}\int_{t}^{t+T_1}\|u(\tau)\|^2_{L^\infty(\Omega)} d\tau \le C_u(t,T_1,\Gamma).
\]
Here,
\begin{align*}
    C_u(t,T_1,\Gamma) =& 2\log\left(1+\frac{4\pi^2 \theta_{t,T_1} \Gamma^2}{\ll_1\ll_2}\right) \left(\frac{(1+\lambda_1)\|\bm f\|^2_{L^\infty}}{(\nu\lambda_1)^2} +  e^{(-\lambda_1 \nu)t }\|\bm u_0\|^2_{H^1_p(\Omega)} \right)    \\
    &+ 2\frac{\|\bm\ell\|^2_{\R_2}}{32\pi^2\Gamma^2},    
\end{align*}
provided 
\[
    \theta_{t,\bar T} = 2\frac{\|\vf\|^2_{L^\infty}}{\bar T\nu^3\lambda_1} + \frac1{\bar T}\int_t^{\bar T+t}\frac{\|\vf(s)\|^2_{\vsLt}}{\nu^2}ds + \frac{2 e^{(-\lambda_1 \nu)t}\|\nabla\bm u_0\|^2_{\vsLt}}{\bar T\nu}.
\]
Finally, we can determine $L$ and $h$ from the following inequalities
\begin{align*} \label{eq:hL}
  h L &\le \frac{4\pi^2(1-\alpha)\nu}{\max\{\ell_1,\ell_2\}},\quad \frac{\min_{\Gamma>0}C_u(t,T_1,\Gamma)}{4\alpha\nu}  <  L (1-\delta).
\end{align*}
This completes the proof.
\end{proof}

\subsection{Scaling behavior of the convergence zone}
The observer convergence theorem provides an upper bound of the estimation error, ${\Sigma^{\frac12}_{h,R}}/{\delta}$. We use ``convergence zone'' ($\Upsilon$) to refer to the upper bound. Here, we show that for a certain class of noise processes, $\bm{\eta}$, increasing the size of $\Omega_j$ improves the upper bound of the convergence zone. 

For a better demonstration, let us simplify the problem setup. The high-resolution data is available at a equi-distance mesh with the grid size of $\delta_x$. The size of the subdomain, $\Omega_j$, is set to satisfy $h=h_1=h_2$, and $\ell_1/h = n_1$ and $\ell_2/h=n_2$ for $n_1,n_2 \in \mathbb{N}$. Furthermore, we set $h = c \delta_x$ for $c \in \mathbb{N}$, so the length of the side of $\Omega_j$ is evenly divisible by $\delta_x$. Let $M$ ad $N$, respectively, denote the total number of grid points, $M = |\Omega|/\delta_x^2 = m_1 m_2$, and the total number of the sub-domains, $N=|\Omega|/|\Omega_j|$, which implies $N=M/c^2$. We call $c$ the \emph{compression ratio}, as it defines the ratio between the high-resolution and the low-resolution data. In order to avoid formal complications of rigorously introducing generalized random vector fields with bounded second moments, we consider a class $\mathcal{N}$ of noise processes $\bm{\eta}=(\eta_1,\eta_2)^\top$ of the following form: if $\bm{\eta}\in\mathcal{N}$ then there exists $\delta_x$ and a uniform partition of $\Omega_j$ by squares $\Omega_{j_{lm}}$ of area $\delta_x^2$ such that for $k=1,2$: $$
\eta_{k}=\sum_{j=1}^N\sum_{l,m=1}^c \eta^{(k)}_{j_{lm}} \xi(\Omega_{j_{lm}},\cdot),   E[\eta^{(k)}_{j_{lm}}\eta^{(k)}_{j_{l'm'}}] = \sigma^2\delta_{ll'}\delta_{mm'}, 
$$ i.e.,  each component of the vector-valued noise $\bm{\eta}$ is a linear combination of indicators $\xi(\Omega_{j_{lm}},\cdot)$ with independent random coefficients $\eta^{(k)}_{j_{lm}}$. Here $\delta_{ll'}$ is Kronecker delta and $\sigma>0$ is the variance.

\begin{cor}\label{c:scaling}
    Let the assumptions of the observer convergence theorem hold, and suppose in addition that $\bm{\eta}\in\mathcal{N}$ and that $R_{1,j}=R_{2,j}\equiv R, \forall j \in \{1,\cdots,N\}$. Then the upper bound of the expected estimation error can be written as
    \begin{equation} \label{eq:scaling}
    E_{\bm{\eta}}\frac{1}{|\Omega|^{1/2}} \|\bm{u}(\cdot,t)-\bm{z}(\cdot,t) \|_{L^2(\Omega)} \le \frac{\sqrt{2}}{\delta}\frac{\sigma}{c}.
    \end{equation}   
 \end{cor}
We provide a demonstration of this corollary for the convenience of the reader. Recall the definition of $R_{k,j}$ from the assumption \eqref{eq:noise_bound} and denote the upper-bound of the convergence zone \eqref{eq:zone2} by  
\begin{equation} \label{eqn:error}
\Upsilon =  \delta^{-1}\Sigma_{h,R}^{\frac{1}{2}},\quad \Sigma_{h,R} = 2h^2\max_k \max_j R^{-1}_{k,j}.
\end{equation}
For every $\bm{\eta}\in\mathcal{N}$ we get: 
\begin{equation*}
E\left[ \frac{1}{h^2}\int_{\Omega_j}\eta_kd\vec{\bm{x}} \right]^2 = E\left[ \frac{\delta_x^2}{h^2}\sum_{l,m=1}^{c} \eta^{(k)}_{j_{lm}} \right]^2 = \frac{\sigma^2}{c^2},
\end{equation*}
where we used that $|\Omega_j| = h^2 = c^2 \delta_x ^2$. By assumption \(
R_{k,j}=R,~\text{for}~\forall k \in \{1,2\}~\text{and}~\forall j \in \{1,\cdots,N\}.
\) Hence, \eqref{eq:noise_bound} yields: 
\begin{equation}
N\frac{\sigma^2}{c^2} R = M \frac{\sigma^2}{c^4} R \le 1,
\end{equation}
which indicates the choice of $R$ for which every $\bm{\eta}\in\mathcal{N}$ verifies \eqref{eq:noise_bound} on average: 
\begin{equation}
R_{max} = \frac{c^2}{N \sigma^2} = \frac{c^4}{M \sigma^2}.
\end{equation}
Then, the expected upper bound for the zone of convergence becomes,
\begin{equation}
\Upsilon = \sqrt{2} h M^{1/2} \sigma (\delta c^2)^{-1} = \sqrt{2} |\Omega|^{1/2} \sigma (\delta c)^{-1}.
\end{equation}
Finally, \eqref{eq:scaling} follows from \eqref{eq:zone2}. Hence, the expected estimation error is inversely proportional to the compression ratio, $c$, and proportional to the variance of the noise provided $\bm{\eta}\in\mathcal{N}$. Note that \eqref{eq:scaling} is independent of $\delta_x$ and, thus, it holds for any noise $\bm{\eta}\in\mathcal{N}$.  

\section*{Numerical setup}

Here, we describe the numerical setup of the experiments. To mimic a realistic setting, the observations are generated with a different solver than what is used for the implementation of the Luenberger observer. 

\subsection*{Computational framework}

We obtain the ground-truth velocity, $\bm{u}(\bm{x},t)$, that is later corrupted with noise and projected onto a lower resolution grid, using the pseudo-spectral method available through JAX-CFD~\cite{Dresdner2022-Spectral-ML}. We, therefore, use a vorticity formulation, which avoids the need to separately
enforce the incompressibility condition $\nabla \cdot v = 0$. We split the linear and nonlinear terms of the
equation into implicit and explicit terms, respectively, and used a Crank-Nicolson time stepping scheme with
fourth-order Runge-Kutta, as described in~\cite{Dresdner2022-Spectral-ML}. The solver is implemented in JAX~\cite{jax2018github} which supports reverse-mode automatic differentiation.~The Kolmogorov flow simulated with JAX-CFD is also reported in~\cite{kochkov2021machine}, but the authors use finite volume method (FVM) on a staggered grid instead. 

To estimate the ground truth, $\bm{z}(\bm{x},t)$, we use a Navier-Stokes solver Oasis (Optimized And StrIpped Solver)~\cite{mortensen2015oasis} written in the Python interface to FEniCS~\cite{logg2012automated}. FEniCS is a popular open-source computing platform for solving partial differential equations (PDEs) with the finite element method (FEM). The transient solver uses the fractional step algorithm~\cite{kim1985application} for the discretization of the Navier-Stokes equations. In the experiments, we use an optimized implementation of a backwards
differencing solver with a pressure correction scheme in rotational form~\cite{guermond2006overview}, referred to as BDFPC\_FAST in Oasis. The linear systems are solved using Krylov methods with a DOLFIN/FEniCS wrapper of PETSc~\cite{petsc-web-page}. Both velocity and pressure are computed using continuous Galerkin with quadratic Lagrange elements.

\subsection*{Experiment configurations}

The divergence-free initial velocity fields are generated with appropriate spectral filtering. The conditions are sampled from the log-normal distribution with the maximum amplitude of the initial velocity field set to 2 and the peak wavenumber 16. The sinusoidal Kolmogorov forcing is applied with a mode 10 and amplitude 1.0;
\[
f_i(\bm{x},t) = \mathrm{sin}(10x_2) ~\text{for}~ i = 1,2.
\]
In other words, $\bm{f}$ is constant in the $x_1$-direction and varies in the $x_2$-direction.

The time-step is set to $\delta t=0.001$ and the data is generated until time $t=15$. As described in the theoretical section of the article, we introduce $c$, the compression ratio, which is defined as $c=\sqrt{M/N}$, where $M$ and $N$, respectively, denote the total number of grid points and the total number of the non-overlapping subdomains that the high-resolution grid is divided into. The computational domain of $2 \pi \times 2\pi$ is represented by $M=512 \times 512$ grid cells for high-resolution and, through average pooling over subdomains, converted to lower-resolution of $N=512 / c \times 512 /c $ grid cells  The setting of experiments is summarized for a range of viscosities, $\nu$, different compression ratios, $c$, and noise levels $\zeta$ in the Table~\ref{experiments-table}. The variance of the observation noise is defined as $\sigma^2 = \zeta^2\times\text{TKE}$, in which TKE denotes the turbulent kinetic energy. All the simulation files needed to reproduce the experiments are available in~\cite{our_software}. The code for data generation with JAX-CFD is based on the script made available in Google Colab~\cite{google_colab}

\begin{table}[!h]
    \begin{tabular}{|c | c|} 
  \hline
  \multicolumn{2}{|c|}{Settings} \\
  \hline
  viscosity, $\nu$ & 0.0015, 0.003, 0.0045, 0.006, 0.0075, 0.009\\
  noise level, $\zeta$ & 0, 0.1, 0.2, 0.4, 0.6, 0.8, 1.6 \\
  compression, $c$ & 1, 2, 4, 8, 16, 32, 64 \\
  number of subdomains, $\sqrt{N}$ & 512, 256, 128, 64, 32, 16, 8 \\
  \hline
    \end{tabular}
\caption{\label{experiments-table} Parameters for numerical experiments.}
\end{table}  

\subsection*{Reconstruction errors}

\begin{figure}
    \centering
    \includegraphics[width=0.32\textwidth]{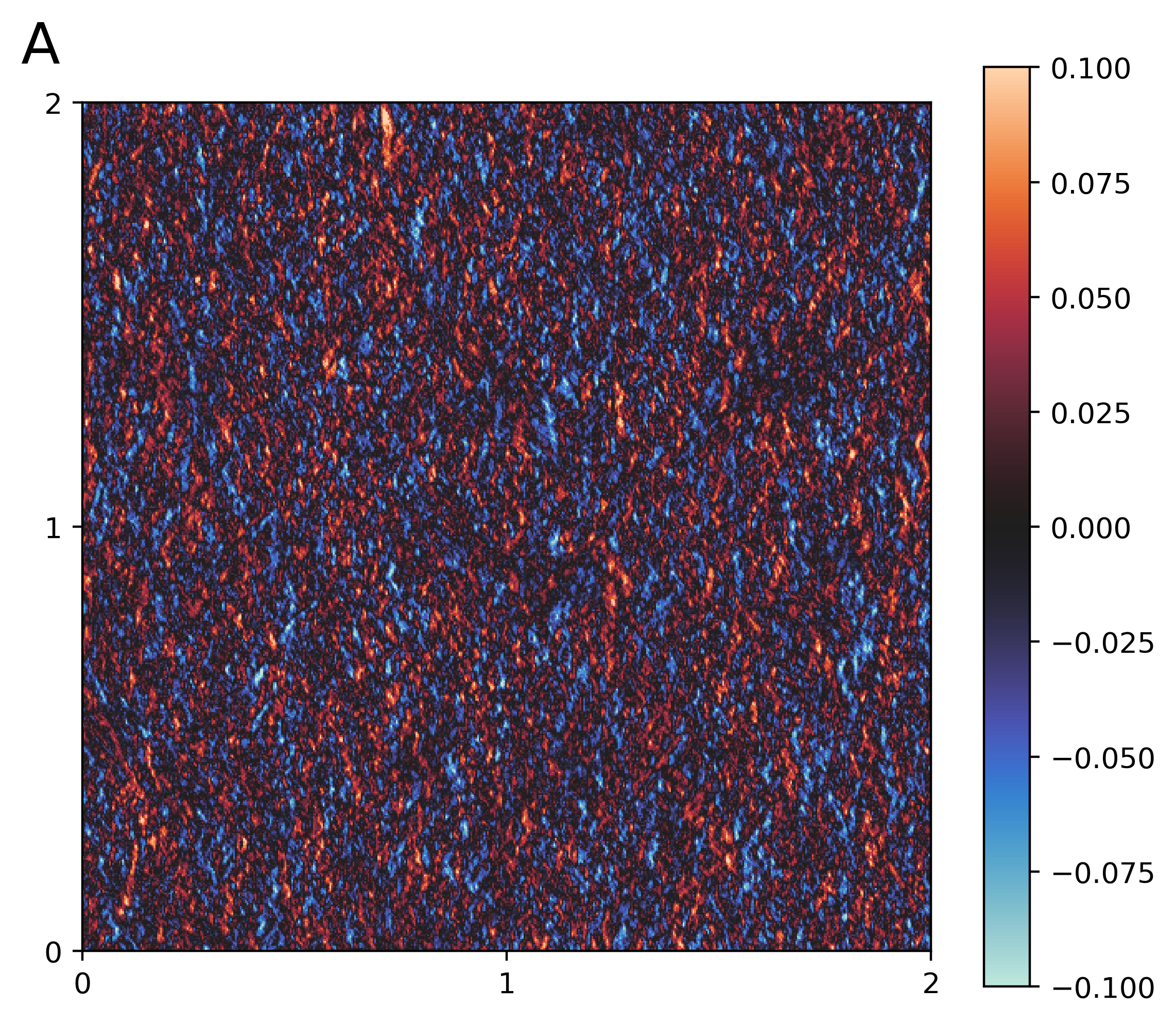}
    \includegraphics[width=0.32\textwidth]{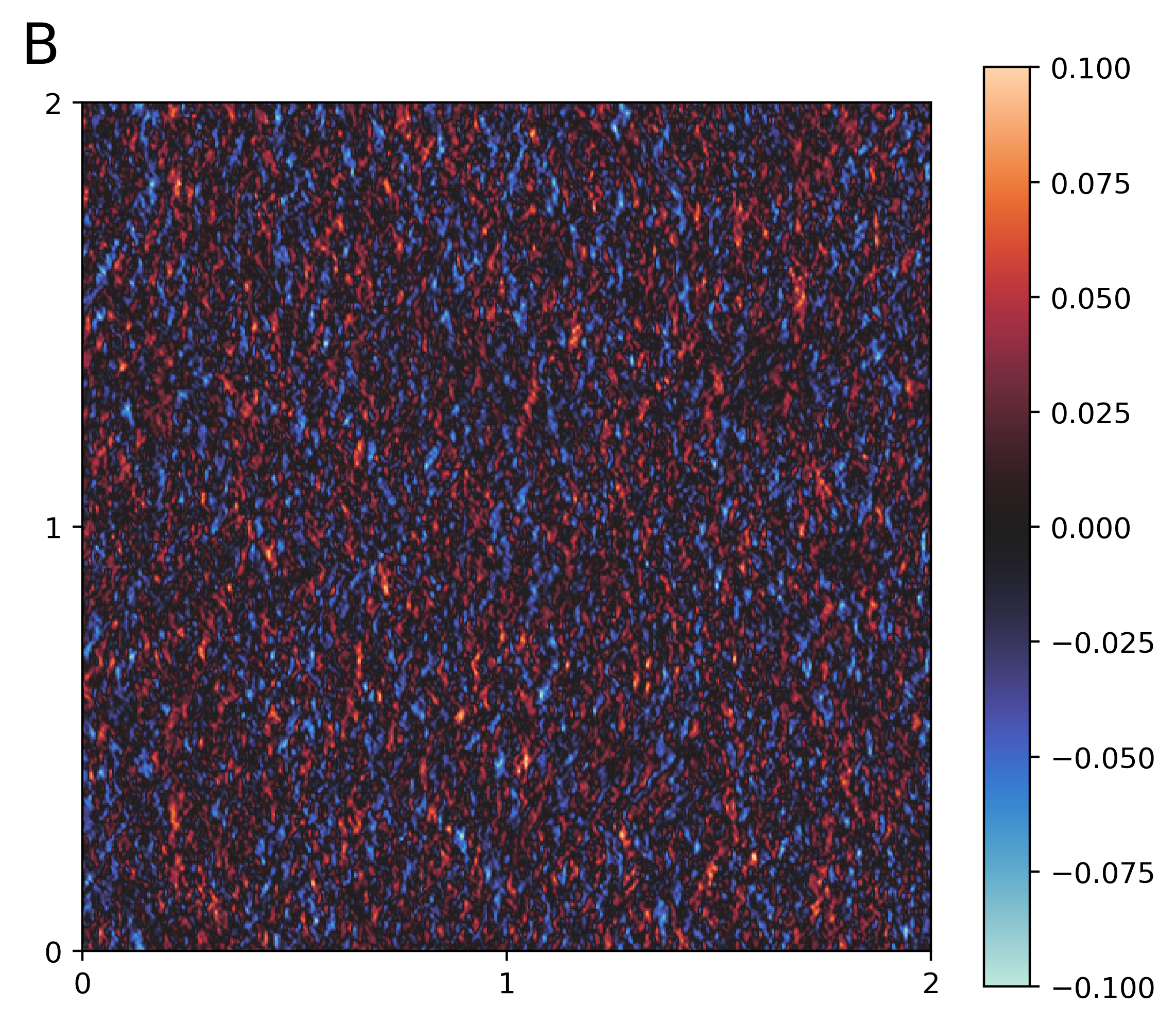}
    \includegraphics[width=0.32\textwidth]{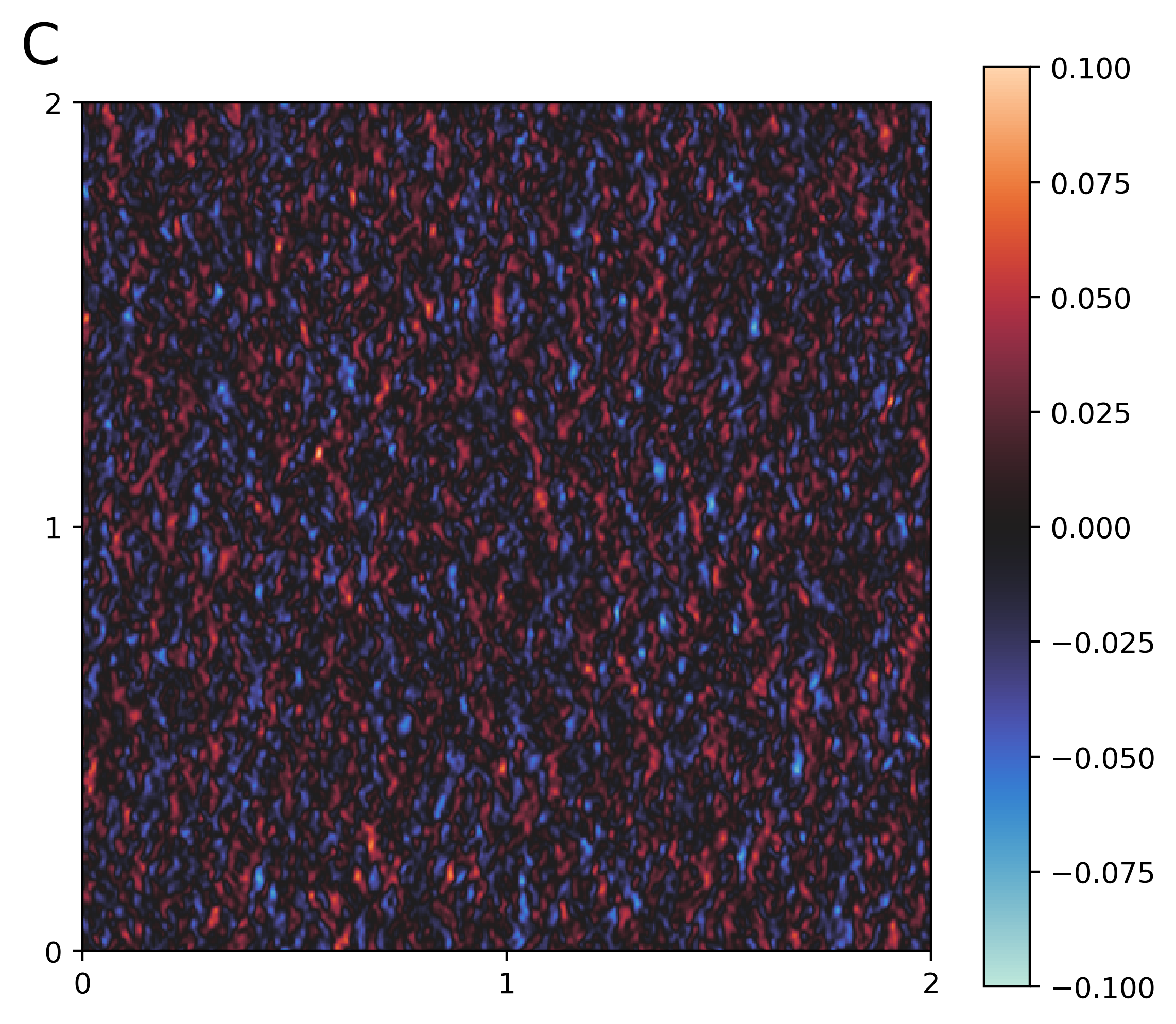}\\
    \includegraphics[width=0.32\textwidth]{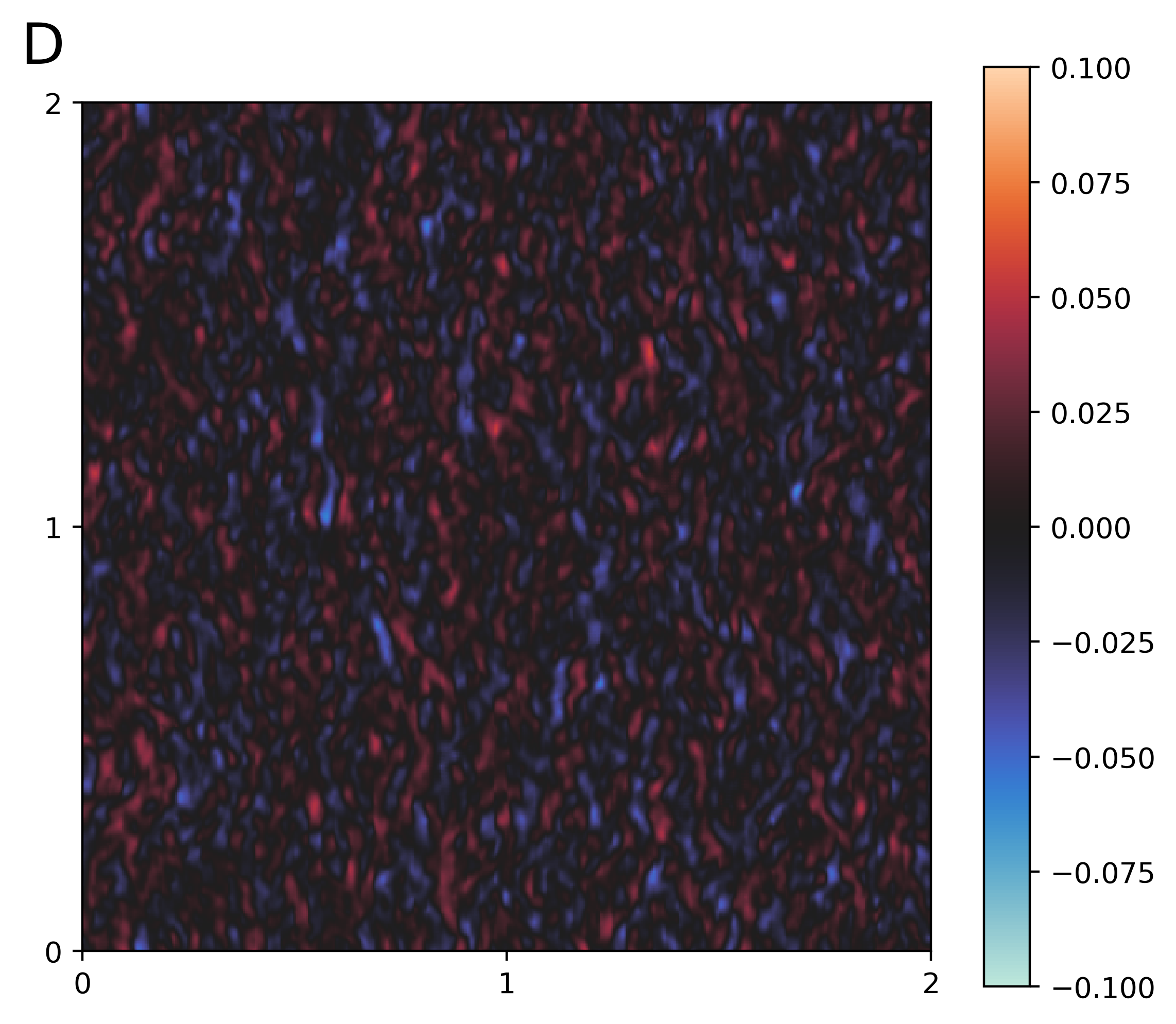}
    \includegraphics[width=0.32\textwidth]{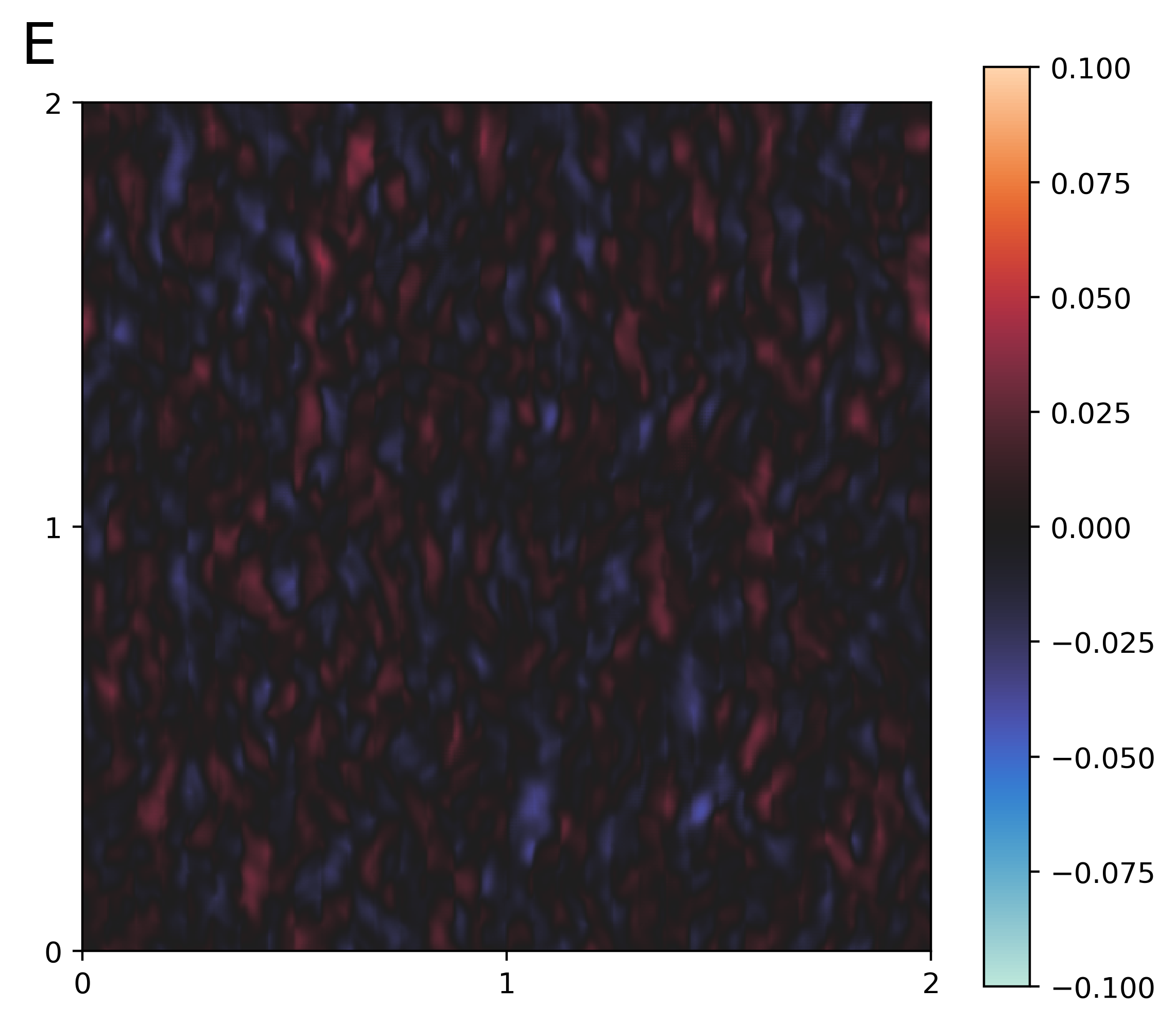}
    \includegraphics[width=0.32\textwidth]{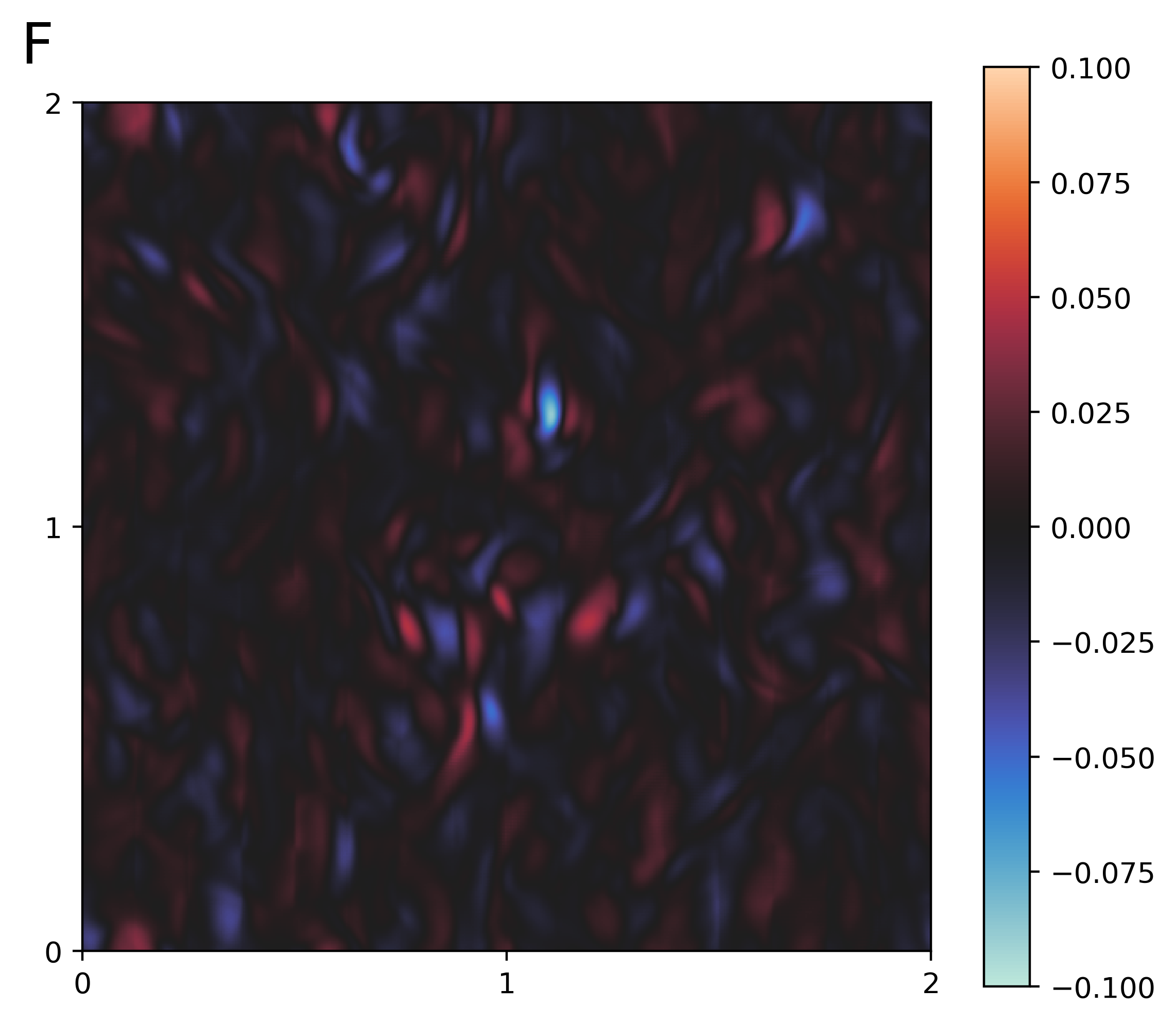}\\
    \caption{Estimation error of $\bm{u}$ in the $x_1$ direction, $e_1(\bm{x},t) = u_1(\bm{x},t)-z_1(\bm{x},t)$ at $t=8$ for different values of the compression ratio ($c$): (A) $c=1$, (B) $c=2$, (C) $c=4$, (D) $c=8$, (E) $c=16$, and (F) $c=32$. The viscosity is $\nu=0.003$ and the noise level is $\zeta = 1.6$.}
    \label{fig:residual}
\end{figure}

The estimation errors, $\bm{e} = \bm{u}-\bm{z}$, for different values of the compression ratio ($c$) are shown in Fig. \ref{fig:residual}. It is shown that, for smaller $c$, $\bm{e}(\bm{x},t)$ almost looks like white noise, indicating the observation noise strongly corrupts the reconstruction. As $c$ becomes larger, in general, the magnitude of $e$ is also attenuated. The maximum deviation, \emph{i.e.}, $l_\infty$ norm of $e_1$, decreases from 14.5 for $c=1$ to 4.9 for $c=16$. For $c=32$, the error is larger than that of $c=16$. This is because, as shown in Fig. 2 B, for $c=32$, the convergence of the observer becomes very slow; the observer achieves a stationary error after $t=10$.

\bibliography{our_bib}